\documentclass[final,2p,times,twocolumn]{elsarticle}

\usepackage{amsmath}
 \usepackage{multirow}
 \usepackage[LGRgreek]{mathastext}
 \usepackage{hyperref}
 \usepackage{rotating}
 \usepackage{graphicx}
 \hypersetup{colorlinks,linkcolor={blue},citecolor={blue},urlcolor={red}}
 \usepackage[final]{changes}
 \colorlet{Changes@Color}{red}
 
\journal{Icarus}
\biboptions{authoryear}

\begin{document}
\begin{frontmatter}



\title{Expected Precision of Europa Clipper Gravity Measurements}


\author[epss]{Ashok K. Verma}
\author[epss,aa]{Jean-Luc Margot}
\address[epss]{Department of Earth, Planetary, and Space Sciences, University of California, Los Angeles, CA 90095, USA}
\address[aa]{Department of Physics and Astronomy, University of California, Los Angeles, CA 90095, USA}

\begin{abstract}
The primary gravity science objective of NASA's Clipper mission to
Europa is to confirm the presence or absence of a global subsurface
ocean beneath Europa's Icy crust.  Gravity field measurements obtained
with a radio science investigation can reveal much about Europa's
interior structure. Here, we conduct extensive simulations of the
radio science measurements with the anticipated spacecraft trajectory
and attitude (17F12v2) and assets on the spacecraft and the ground,
including antenna orientations and beam patterns, transmitter
characteristics, and receiver noise figures.  In addition to two-way
Doppler measurements, we also include radar altimeter crossover range
measurements.  We concentrate on $\pm$2 hour intervals centered on the
closest approach of each of the 46 flybys.  Our covariance analyses
reveal the precision with which the tidal Love number $k_2$,
second-degree gravity coefficients $\bar{C}_{20}$ and $\bar{C}_{22}$,
and higher-order gravity coefficients can be determined.  The results
depend on the Deep Space Network (DSN) assets that are deployed to
track the spacecraft. We find that some DSN allocations are sufficient
to conclusively confirm the presence or absence of a global
ocean. Given adequate crossover range performance, it is also possible
to evaluate whether the ice shell is hydrostatic.
\end{abstract}


\end{frontmatter}


\section{Introduction}
 \label{intro}
The spacecraft mission design process at NASA relies on design
requirements that flow from measurement requirements, which themselves
flow from science objectives.  The Europa Clipper mission has a set of
compelling science objectives~\citep[e.g.,][]{Pappalardo17} that
emerged out of strategic planning
documents~\citep[e.g.,][]{nrc99,decadal11} and other studies.  Here we
investigate some of the measurement requirements that may be needed to
enable a gravity science investigation.  Gravity science experiments
provide powerful data for investigating the physical state of
planetary bodies.  Examples include mapping the gravity field,
estimating the rotational state, and probing the internal structure of
Mercury \citep[e.g.,][]{Smith12,maza14,Verma16}, 
Venus \citep[e.g.,][]{sjog97,kono99}, 
Mars \citep[e.g.,][]{Smith99,Konopliv11}, 
and Titan \citep[e.g.,][]{Iess10}.

In 2015, NASA appointed a Gravity Science Working Group (GSWG) to help
refine science objectives for the Europa Clipper mission (then known
as the Europa Multiple Flyby Mission).  NASA's charge to the GSWG
included the following statement: ``The GSWG will define and recommend
to the science team science investigations related to understanding
the response of the satellite to gravity, specifically, but not
limited to, understanding the tidal distortion of Europa, its internal
structure, precession, and moments of inertia.''  The GSWG produced a
report~\citep{gswg} that specifies the precision with which certain
quantities must be measured in order to meet specific science
objectives (Table~\ref{tab-gswg}).  The GSWG focused primarily on
measurements that pertain to the ice shell and the presence of an
ocean.

\begin{table*}[!h]
\centering
\caption{A subset of possible measurement objectives for a Europa
  Clipper gravity science investigation \citep{gswg}. The rightmost
  column specifies the one-standard-deviation precision with which
  geophysical parameters must be measured in order to meet gravity
  science objectives.  The GSWG recommended multiplying formal
  uncertainties of fitted parameters by a factor of two to arrive at
  realistic one-standard-devation uncertainties -- see
  Section~\ref{sec-covariance}.  The spherical harmonic coefficients
  in the representation of the gravity field are 4$\pi$-normalized.
  In this work, we focus on the first three objectives.}
\begin{tabular*}{\textwidth}{l@{\extracolsep{\fill}}ll}
\hline
Objective                                     & Quantity                & Required precision \\
\hline
Confirm the presence of an ocean              & Tidal Love number $k$       & $k_2$$\textless$0.06  \\
Verify whether ice shell is hydrostatic       & Gravitational harmonics & $\bar{C}_{20}$$\textless$8e-6 and $\bar{C}_{22}$$\textless$9e-6 \\
Measure shell thickness (to $\pm$20$\%$)  & Gravitational harmonics & $\bar{C}_{30}$$\textless$4e-7 and $\bar{C}_{40}$$\textless$4e-7 \\
Confirm the presence of an ocean              & Tidal Love number $h$       & $h_2$$\textless$0.3  \\
Confirm the presence of an ocean              & Obliquity                      & $\theta$$\textless$0.01$^{\circ}$  \\
Measure elastic shell thickness (to $\pm$10 km) & Tidal Love numbers & $k_2$$\textless$0.015 and $h_2$$\textless$0.015 \\
Confirm ice shell is decoupled from interior &  Amplitude of longitude libration & $\textless$50 m at tidal period \\
\hline
\label{tab-gswg}
\end{tabular*}
\end{table*}

One of the primary objectives of a mission to Europa is to confirm the
presence of a global ocean.  A gravity science investigation can
address this objective in a number of ways~\citep{gswg}.  Here, we
focus on  measurements of the tidal Love number $k_2$. 
An alternate approach consists of measuring the tidal Love number
$h_2$, as examined by \citet{Steinbruegge17}.  Calculations by
\cite{Moore2000} indicate that $k_2$ is expected to range from 0.14 to
0.26, depending on the thickness and strength of the ice shell, if a
global ocean is present underneath the ice shell.  In contrast, $k_2$
is expected to be less than 0.015 if there is no global ocean.
Therefore, a measurement of $k_2$ is sufficient to test the global
ocean hypothesis~\citep{park2011,park2015,maza15}, provided that the
uncertainties do not exceed the 0.06 level recommended by the GSWG.

Another important objective of a gravity science investigation is to
confirm whether the ice shell is in hydrostatic equilibrium.
Galileo-based estimates of second-degree gravity coefficients rely on
the assumption of hydrostatic equilibrium \citep{Anderson98}, but it
is unclear whether hydrostatic equilibrium applies.  It is possible to
test the hydrostatic equilibrium hypothesis by measuring the
second-degree gravitational harmonic coefficients, $\bar{C}_{20}$ and
$\bar{C}_{22}$, to the level prescribed by the GSWG
(Table~\ref{tab-gswg}).  Trajectories being designed for the Clipper
mission offer promising prospects for measuring these quantities.

In section \ref{refTraj}, we provide an overview of the anticipated
Clipper trajectory.  In section \ref{sec-msr}, we review measurements,
uncertainties, and model assumptions.  Our dynamical model, solution
strategy, and estimated parameters are discussed in section
\ref{method}. In Section \ref{res}, we discuss our covariance analysis
results.  Our conclusions are provided in section \ref{conclusion}.

\section{Spacecraft trajectory and attitude}
\label{refTraj}

\begin{figure*}
\centering
\noindent
\includegraphics[width=30pc]{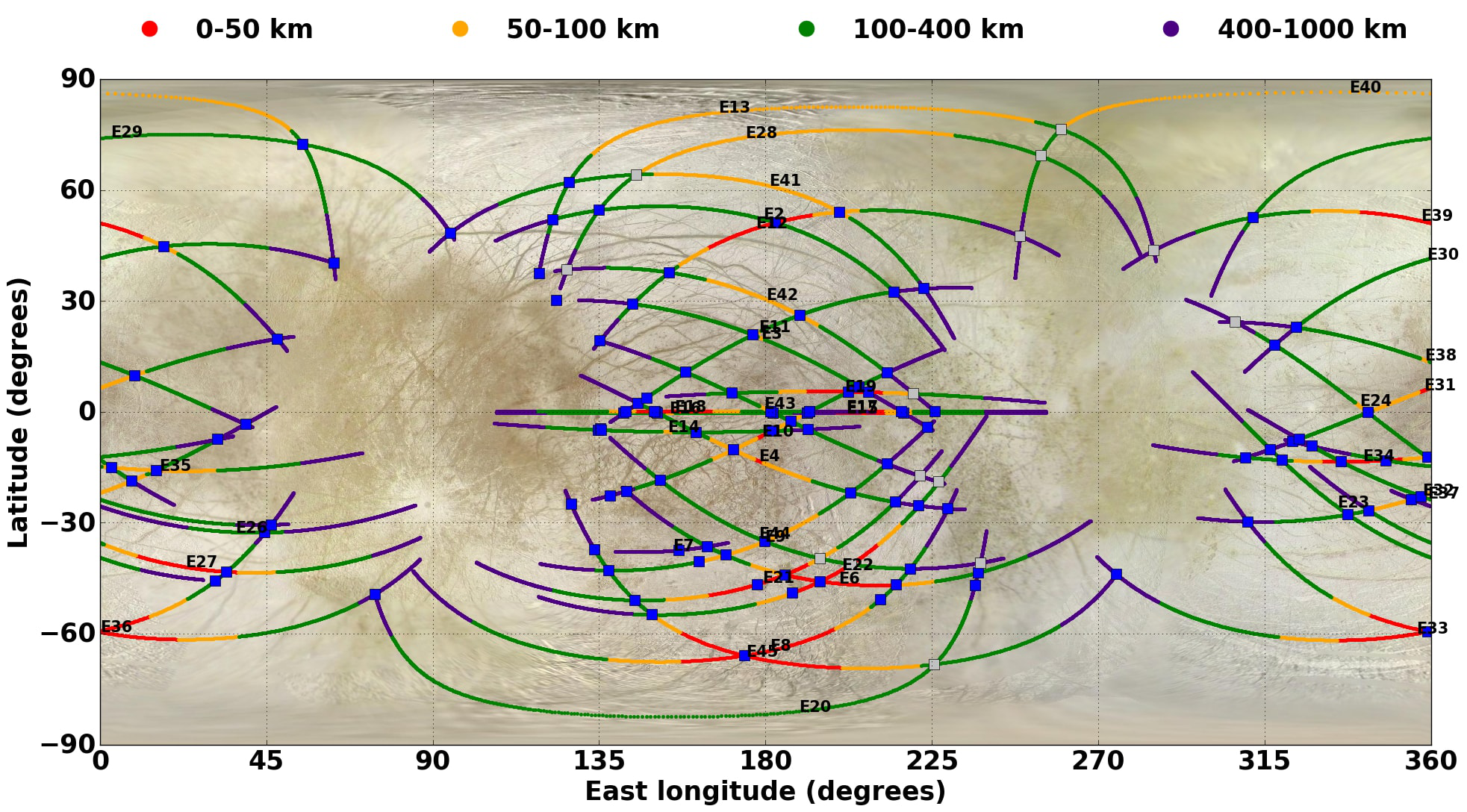}
\caption{Ground tracks (solid lines) and crossover locations (squares)
  corresponding to trajectory 17F12v2.  Ground tracks are color-coded
  by altitude and crossovers are color-coded in blue (when Europa's
  surface is illuminated by the Sun) or silver (when the surface is in
  darkness).  Only crossovers that occur when both altitudes are below
  1000 km are shown.}
\label{nomTraj}  
\end{figure*}

Europa Clipper will orbit Jupiter and execute repeated close flybys of
Europa, Ganymede, and Callisto with science observations at Europa and
gravitational assists at Ganymede and Callisto \citep{Lam15}.  To
achieve the science goals of the mission, Clipper trajectories are
designed to obtain globally distributed regional coverage of Europa
with multiple low-altitude flybys \citep{Pappalardo17}.  The current
trajectory, named 17F12v2, includes 46 flybys with altitudes as low as
25~km (Figure \ref{nomTraj}) and 126 crossovers below 1000 km
altitude.  Crossovers are locations where two ground tracks intersect
and where altimetric measurements are particularly valuable.

We examined the suitability of trajectories 15F10, 16F11, and 17F12v2
for gravity science investigations, with a particular emphasis on
17F12v2.  \added{All of these trajectories were designed to obtain
  globally distributed regional coverage of Europa with 42, 43, and 46
  flybys, respectively (Table \ref{TrjTable}).}

\begin{table*}[!h]
\centering
\caption{Characteristics of trajectories considered in this work: number of flybys according to closest approach altitude and number of illuminated crossovers with closest approach altitude below 1000 km.}
\label{TrjTable}
\begin{tabular*}{\textwidth}{l@{\extracolsep{\fill}}ccccc}
\hline
\multirow{2}{*}{Trajectory} & \multicolumn{4}{c}{Flybys} & \multirow{2}{*}{\begin{tabular}[c]{@{}c@{}}Illuminated crossovers\\  
\textless 1000 km\end{tabular}} \\ \cline{2-5}
& \textless 50 km & 50--100 km & 100--1000 km & \textgreater 1000 km &  \\
\hline
15F10                       & 22              & 15        & 1           & 4                    & 88       \\
16F11                       & 26              & 13        & 2           & 2                    & 106      \\
17F12v2                     & 25              & 16        & 2           & 3                    & 112       \\
\hline                                                                                              
\end{tabular*}
\end{table*}

An important consideration for a gravity science investigation is the
distribution of sub-spacecraft latitudes when the spacecraft is at
closest approach.  Trajectory 17F12v2 provides an adequate
distribution for gravity science purposes (Table
\ref{nomTrajTable}). Details about the spacecraft's anticipated
trajectory and orientation in space (attitude) are available \added{at
  ftp://naif.jpl.nasa.gov/pub/naif/EUROPACLIPPER} in the form of SPICE
kernels~\citep{acto17}.

\begin{table*}[!h]
\centering
\caption{Definition of latitude regions and latitudinal distribution of
  flybys at the epochs of closest approach.  The range of closest approach altitudes is also shown. } 
\begin{tabular*}{\textwidth}{l@{\extracolsep{\fill}}  c c c}
\hline  
Europa region         &  Latitude range                     & Altitude range       &   Number of flybys   \\ 
\hline
High latitude north  &              90$^\circ$ $\textendash$  45$^\circ$ &   25 km $\textendash$ 1442 km        & 9  \\
Mid latitude north   &              45$^\circ$ $\textendash$  15$^\circ$ &    50 km $\textendash$ 100 km      &  4  \\
Low latitude         &              15$^\circ$ $\textendash$ -15$^\circ$ &    25 km $\textendash$ 100 km     & 13  \\
Mid latitude south   &             -15$^\circ$ $\textendash$ -45$^\circ$ &     25 km $\textendash$ 2554 km    & 12 \\
High latitude south  &             -45$^\circ$ $\textendash$ -90$^\circ$ &    25 km $\textendash$ 100 km     & 8 \\
\hline
\label{nomTrajTable}
\end{tabular*}
\end{table*}

\section{Measurements}
\label{sec-msr}

The gravity science investigation will utilize a radio link between
Earth-based stations and the spacecraft's radio frequency
telecommunications subsystem to provide range and Doppler measurements 
\added{(see Section \ref{dopMsr})} and solve for Clipper's trajectory.  These data will yield
measurements of Europa's gravity field and tidal response.  The
investigation will also rely on \added{spacecraft-to-Europa} ranging data from the 
Radar for Europa Assessment and Sounding: Ocean to Near-surface (REASON)
instrument~\citep{blan14}. 
Analysis of REASON data may be enhanced with digital elevation models
obtained by stereo imaging from the Europa Imaging System
(EIS)~\citep{turt16}.  The primary observables will be two-way,
coherent Doppler measurements (section \ref{dopMsr}) and radar range
measurements (section \ref{croMsr}).  Our assumptions about the radio
link and the Doppler and crossover measurements are summarized in
Table~\ref{tab-assumptions} and explained in detail below.

\subsection{Radio link}
In order to meet mission requirements, it is anticipated that Clipper
will carry at least three fan-beam (medium-gain) antennas and two
wide-beam (low-gain) antennas.  Nominal antenna gain patterns for
these antennas were provided by Peter Ilott (pers.\ comm.) and Avinash
Sharma (pers.\ comm.).  \added{We used these gain patterns in conjunction with 
the spacecraft attitude to compute the signal to noise ratio of the radio link.}
Clipper will carry a 20 W transmitter operating in the X band.  Transmitter parameters 
were provided byDipak Srinivasan (pers.\ comm.). Ground stations are expected to be
primarily 34~m or 70~m antennas of the Deep Space Network (DSN)
equipped with low-noise receivers.  Nominal DSN system temperatures were 
provided by Ryan Park (pers.\ comm.).
\replaced{Typical spacecraft radio links are established in a closed-loop mode
with a signal to noise ratio of 7 dB-Hz or above.}  
{A typical spacecraft radio link operates 
with a signal to noise ratio of 7 dB-Hz or above.}
\replaced{However, it may be possible to use an open-loop receiver and obtain radio science data with a signal to noise ratio of 4 dB-Hz or less~\citep{dsn}.}{However, it may be possible to  establish a radio link with 4 dB-Hz.}
The DSN has the capability to array two or three
34~m antennas to improve the radio link budget.

\subsection{Doppler measurements}
\label{dopMsr}

Gravity science investigations rely primarily on two-way Doppler shift
measurements between spacecraft and Earth-based antennas.  These
measurements yield the velocity of the spacecraft along the observer
line-of-sight (LOS). Because the Doppler shift measures a component
of the spacecraft velocity that is affected by the gravitational
field, the radio tracking of spacecraft can produce detailed
information about the distribution of mass in planetary bodies. To
first order,
\begin{equation}
\label{dopEq}
f_D \simeq 2f_T\frac{V_r}{c},
\end{equation}
where $f_D$ is the Doppler shift, $f_T$ is the transmitted frequency,
V$_r$ is the LOS component of the relative velocity between spacecraft
and observer, and $c$ is the speed of light.

\begin{figure*}
\centering
\noindent
\includegraphics[width=30pc]{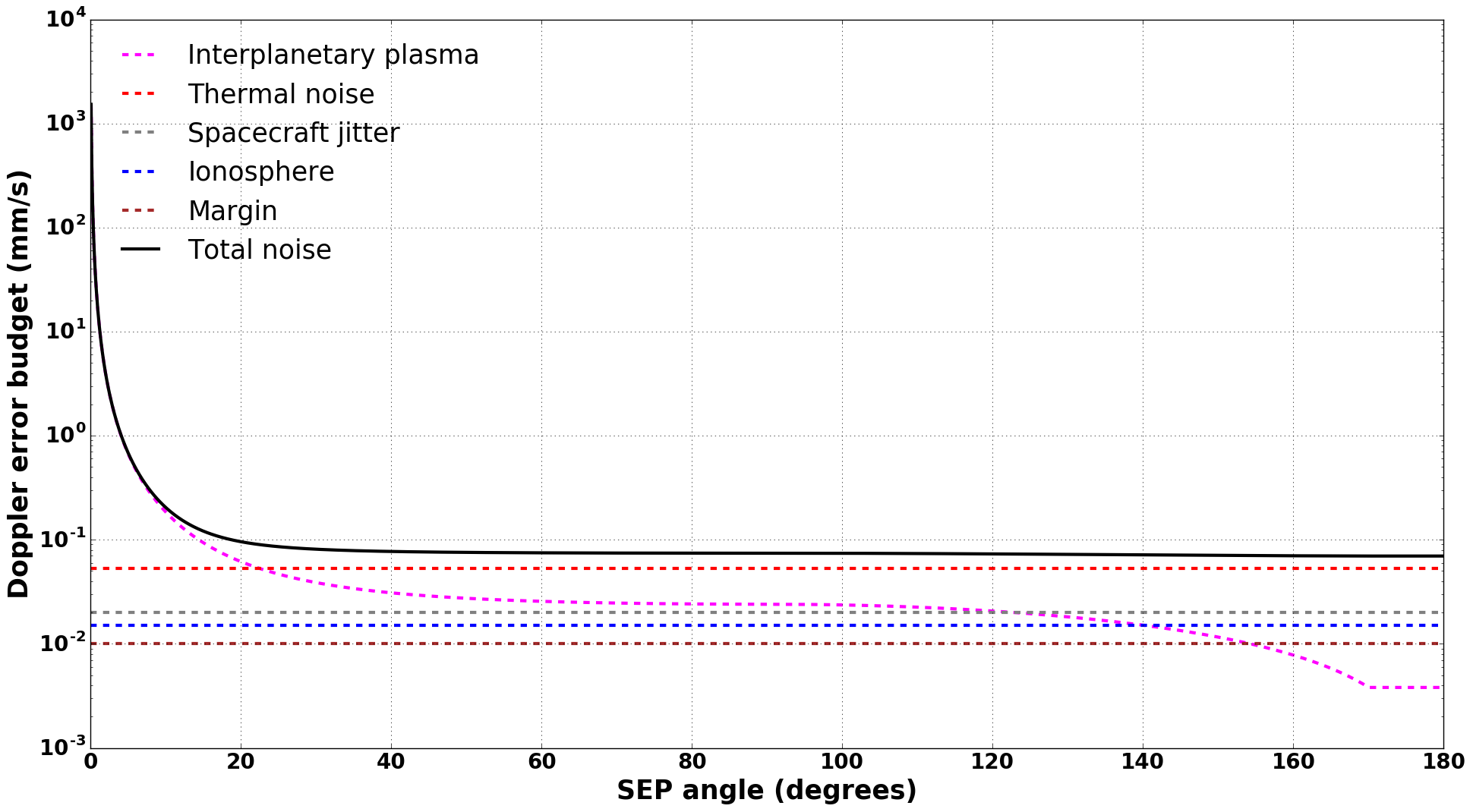}
\caption{Graphical representation of the Doppler error budget adopted
  in this work (equation (\ref{eq-dopSigma})), showing Doppler
  uncertainties as a function of SEP angle.  The model is appropriate
  for two-way Doppler measurements at X-band and 60 s integration
  time.}
\label{fig-dopSigma}  
\end{figure*}

In this study, we assumed that the spacecraft telecom subsystem
receives an X-band signal with a carrier frequency of $\sim$7.2 GHz
from a DSN ground station and coherently transmits this signal back to
the DSN with a carrier frequency of $\sim$8.4 GHz.  The uncertainties
of the Doppler measurements depend on a number of factors that include
fluctuations in the ionospheric and solar wind plasmas, variations in
the water content in the troposphere, as well as instrumental noise
\citep{Asmar05}.  However, at small Sun-Earth-Probe (SEP) angles, the
\added{interplanetary} plasma noise dominates.
We modeled the Doppler uncertainties in a 60 s integration time as:
\begin{equation}
  \label{eq-dopSigma}
  \sigma_D = \sqrt{ \sigma_{plasma}^2 + \sigma_{other}^2} + 0.01 \ mm/s,
\end{equation}
\replaced{where $\sigma_{plasma}$ represents the noise due to interplanetary
plasma according to the model of \cite{Iess2012} and $\sigma_{other}$ is
a Europa Project estimate of the noise contribution due to other sources, 
including thermal noise (0.053 mm/s), spacecraft jitter (0.020 mm/s), and ionosphere (0.015 mm/s).  
The last term (0.01 mm/s) represents a margin added to the noise model (Figure \ref{fig-dopSigma}).  
See \citet{Asmar05} for a detailed review of noise sources in radio science experiments.} 
{where $\sigma_{plasma}$ represents the noise due to interplanetary
plasma according to the model of Iess et al. (2012), $\sigma_{other}$ is
a Europa Project estimate (0.0596 mm/s) of the noise contribution due
to other sources, including media, DSN clocks, and spacecraft and DSN
instruments, and 0.01 mm/s is a small margin added to the noise 
model (Figure \ref{fig-dopSigma}).}

\subsection{Crossover measurements} 
\label{croMsr}

Improvements to the quality of a spacecraft's orbit determination can
be obtained when ranging measurements to a body of known shape and
surface of known topography are available.  Even when the shape varies
with time and when the topography is unknown, as is the case for
Europa, it is still possible to benefit from altimetry at so-called
crossover points.  Crossovers points are the geographic locations on
the surface where two ground tracks intersect (Figure \ref{nomTraj}).
Ranges obtained at the same location on two separate tracks can be
subtracted to yield a crossover height difference:
\begin{equation}
\label{croEq}
\Delta h(t_1, t_2) = h(t_1)-h(t_2),
\end{equation}
where $h(t)$ is the altimeter height measurement at time $t$.  These
crossover measurements eliminate the uncertainty introduced by the
unknown local topography and can provide powerful constraints in the
orbit determination process.  The analysis of multiple crossover
measurements obtained throughout Europa's orbital cycle will also be
important in determining the radial amplitude of the tidal signal,
i.e., in estimating the tidal Love number $h_2$
\citep[e.g.,][]{wahr06}.

To simulate crossover measurements, we assumed a nadir-pointed
altimeter and incorporated spacecraft attitude data in our
calculations.  It is anticipated that the REASON instrument will range
to the surface when the spacecraft is below altitudes of 1000~km with
respect to the surface of Europa.  We identified 126 crossovers with
spacecraft altitudes $<$~1000~km in trajectory 17F12v2.  However, 14
of these crossovers were discarded because they occur when the terrain
under the spacecraft is not illuminated by the Sun during one or both
of the encounters.  Without proper illumination, it may not be
possible to produce a Digital Terrain Model (DTM) from stereo analysis
of EIS images.
And without a DTM, the enhanced cross-over analysis approach that
combines REASON and EIS data, as described by \citet{Steinbruegge17},
would not be possible.

We assigned uncertainties to the crossover measurements given by
\begin{equation}
\label{croEq}
\sigma_{t_1,t_2} = \sqrt {(\sigma_{t_1}^2 + \sigma_{t_2}^2)},
\end{equation}
\noindent where $\sigma_{t}$ is the altimeter height measurement
uncertainty at time $t$.  These uncertainties were provided by the
REASON team (Gregor Steinbruegge, pers.\ comm., Sept. 30, 2017) with
heritage from the procedure described by \citet{Steinbruegge17}.
The median and standard deviation of
crossover uncertainties are 4.5 m and 6.5 m, respectively, with
minimum and maximum values of 2.7 and 24.9 m, respectively.

\begin{table*}
\small
\caption{Summary of assumptions regarding observables and radio science system. }
\centering
\begin{tabular*}{\textwidth}{l@{\extracolsep{\fill}}  l l }
\hline
Type        &  Assumption                      &  Notes      \\ 
\hline
                &    X-band uplink          &   reference frequency $\sim$7.2 GHz \\
       &      X-band downlink         &  reference frequency $\sim$8.4 GHz\\
Doppler        &   60 s count time  & integration time for Doppler measurements\\
measurements \hspace{.5cm}       &   up to $\pm$2 h of tracking per flyby  & centered on closest approach epochs\\
                 &   0.07 mm/s $ \le \sigma_D \le$ 0.096 mm/s  & at 20$^{\circ}$ $\le$ SEP $\le$ 180$^{\circ}$ \\
                 &  0.096 mm/s $ \le \sigma_D \le$  0.21 mm/s & at 10$^{\circ}$ $\le$ SEP $\le$ 20$^{\circ}$ \\
\hline
                 & surface illumination by the Sun       & excludes crossovers over non-illuminated terrain\\
        & $<$ 1000 km altitude                  &  spacecraft altitude at epochs of crossovers\\
  Crossover   & range uncertainties:                       &  from REASON team\\
   measurements              & \hspace{5mm} median = 4.5  m & from REASON team\\
                 & \hspace{5mm} standard deviation = 6.5 m & from REASON team\\
   & \hspace{5mm} [min, max] = [2.7, 24.9] m                           & from REASON team\\
\hline
 & +40 dBm transmitter power   & includes all losses due to waveguides and switches \\
 Spacecraft & 3 fan-beam antennas (FBAs)  & orientation per 2c spacecraft model (Version 0.1)\\
 transmitter & 2 low-gain antennas (LGAs)  & orientation per 2c spacecraft model (Version 0.1)\\
  and & LGAs: Gaussian beam pattern & peak at 7.5 dB and width of 42$^\circ$ \\
  antennas & FBAs: fan-beam pattern & $G(\theta,\phi)$ from table lookup and bi-linear interpolation\\
   & No obscuration by spacecraft structures & assumes free path, no scatter losses included \\
 \hline
 & 68.4 dBi antenna gain & for 34 m antenna~\citep{DSN810}\\
 & 31.29 K system temperature & for 34 m antenna (from Ryan Park)\\
  & 74.3 dBi antenna gain & for 70 m antenna~\citep{DSN810}\\ 
   & 26.39 K system temperature & for 70 m antenna (from Ryan Park)\\
 DSN antennas & 1 Hz downlink loop bandwidth & used in noise power calculation \\
 & 2.71 dB two-antenna array gain & assumes typical 0.3 dB combining loss~\citep{DSN810}\\
 & 4.47 dB three-antenna array gain &  assumes typical 0.3 dB combining loss~\citep{DSN810} \\
 & Jupiter radio noise contribution & model from \citet{DSN810}; assumes 152 K black body\\
 & No elevation or horizon mask      & assumes continuous view of Jupiter by DSN assets\\
 \hline
 & Earth-Jupiter distance: & trajectory 17F12v2\\
  & \hspace{5mm} mean = 5.36 au & trajectory 17F12v2\\
Radio link    & \hspace{5mm} median = 5.39  au & trajectory 17F12v2\\
 & \hspace{5mm} [min, max] = [4.36, 6.28] au &  trajectory 17F12v2\\
 & 4 dB-Hz link budget & best-effort basis\\
 \hline                
\label{tab-assumptions}
\end{tabular*}
\end{table*}

\section{Methodology}
\label{method}
Our approach consisted in simulating the precision with which the
Love number $k_2$ and degree-two gravitational harmonics can be
determined with realistic mission scenarios and assumptions about
measurement uncertainties (section~\ref{sec-msr}).

We used version 124 of the Mission Operations and Navigation Toolkit
Environment (MONTE) \citep{Evans16}, an astrodynamics computing
platform that is developed at NASA's Jet Propulsion Laboratory (JPL).
In addition to its uses in trajectory design and spacecraft
navigation, MONTE has been used for a variety of scientific purposes,
including gravity analysis \citep{Verma16}, orbit determination
\citep{Greenberg17}, and sensitivity analysis for tests of general
relativity~\citep{Verma17}.

MONTE's observation model uses \citet{Moyer}'s formulation to compute
Doppler measurements (section \ref{dopMsr}) and a ray-intersect method
to compute altitude measurements (section \ref{croMsr}).  MONTE can
specify arbitrary gravity fields (section \ref{sec-grav}) and spin
states (section \ref{sec-spin}).  We used MONTE's observation model to
compute simulated observables and their partial derivatives with
respect to solve-for parameters (section \ref{sec-params}).
Finally, we used MONTE's tools to perform covariance analyses and quantify the
precision with which geophysical parameters can be determined (section \ref{sec-covariance}).
In the sections below, we provide details about
the gravity and spin state models, solve-for parameters, and
covariance analyses.

\subsection{Representation of Europa's gravity field}
\label{sec-grav}
MONTE represents gravity fields as spherical harmonic expansions \citep[][]{Kaula00}:
\begin{equation}
 \label{potential}
\begin {aligned}
U =  {} & \frac{GM}{r} + \frac{GM}{r} \sum_{l=2}^{\infty}\sum_{m=0}^{l} \left(\frac{R}{r}\right)^l \bar{P}_{l,m} (\sin\phi)  \\
 & \bigg(\bar{C}_{l,m} \cos(m\lambda) + \bar{S}_{l,m} \sin(m\lambda) \bigg),
\end{aligned}
\end{equation}

\noindent where $G$ is the gravitational constant, $M$ is the mass of
Europa, $\bar{P}_{l,m}$ are the normalized associated Legendre
polynomials of degree $l$ and order $m$, $R$ is the reference radius
of Europa (1562.6 km, \citet{arch11}), and $\lambda$, $\phi$, and $r$
are the longitude, latitude, and distance of Clipper from the origin
of the reference frame, which is chosen to coincide with Europa's
center of mass.  $\bar{C}_{l,m}$ and $\bar{S}_{l,m}$ are the
$4\pi$-normalized dimensionless spherical harmonic coefficients.  In
this work, we limited gravity field expansions to degree and order 20.

Jupiter's gravity field induces tides in Europa.  Because of the small
eccentricity of Europa's orbit, the tidal amplitude varies as a
function of time.  MONTE represents the tidal signal by applying
time-varying corrections to the spherical harmonics coefficients
\citep[][p.59]{McCarthy04}:
\begin{equation}
 \label{loveEq}
\begin {aligned} 
\Delta{\bar{C}_{2,m}} - i\Delta{\bar{S}_{2,m}} = {} &  \left(\frac{k_{2,m}}{5}\right)    \left(\frac{M_{J}}{M}\right)  
\left(\frac{R}{r_{EJ}}\right)^{3}   
\bar{P}_{2,m} (\sin\phi_J) e^{-im\lambda_J},
\end {aligned}
\end{equation}
where $\Delta{\bar{C}_{2,m}}$ and $\Delta{\bar{S}_{2,m}}$ are the
time-varying corrections to $\bar{C}_{2,m}$ and $\bar{S}_{2,m}$,
respectively, $k_{2,m}$ is the \deleted{degree-2} Love number \added{for degree 2 and order $m$}, $M_{J}$ is the mass
of Jupiter, $r_{EJ}$ is the distance between Jupiter and Europa, and
$\lambda_J$ and $\phi_J$ are the East longitude and latitude of the
sub-Jupiter point in Europa's body-fixed frame.  \added{Here, we assume $k_{2,0} = k_{2,1} = k_{2,2} = k_{2}$.}

\subsection{Representation of Europa's spin state}
\label{sec-spin}

The orientation of planetary bodies in inertial space can reveal
important insights about interior properties.  Europa's spin state is
not know well.  Its spin period is thought to be closely synchronized
to its orbital period, and its obliquity is thought to be
small~\citep{Bills09}.  Analysis of existing Earth-based radar
measurements is expected to provide a measurement of Europa's spin
axis orientation to arcminute precision \citep{marg13dps}.

The orientation of Europa can be modeled as:
\begin{equation}
 \label{alphadot}
\alpha = \alpha_0 + \dot\alpha\Delta t + \sum_{i} A_i \sin M_i,
\end{equation}
\begin{equation}
 \label{deltadot}
\delta = \delta_0 + \dot\delta\Delta t + \sum_{i} B_i \sin M_i,
\end{equation}
\begin{equation}
 \label{wdot}
W = W_0 + \dot W\Delta t + \sum_{i} C_i \sin M_i,
\end{equation}
where $\alpha$ and $\delta$ are the right ascension and declination of
the spin pole, respectively, $W$ is the orientation of the prime
meridian, $\alpha_0$, $\delta_0$, and $W_0$ are the values at the
J2000 reference epoch, $\dot\alpha$, $\dot\delta$, and $\dot W$ are
the corresponding rates of change, $\Delta t$ is the time since the
reference epoch, and the
A$_i$, B$_i$, C$_i$, and $M_i$ describe Fourier expansions of the
nutation-precession and libration signatures.  In this work, we use
the rotation model that is recommended by the International
Astronomical Union's Working Group on Cartographic Coordinates and
Rotational Elements~\citep{arch11}, which has its origin in
\citet{lies98}'s model.
The values of the coefficients can be found in the current version
(pck00010.tpc) of the planetary constants kernel published by NASA's
Navigation and Ancillary Information Facility (NAIF) \citep{acto17}.

\subsection{Solve-for parameters}
\label{sec-params}

In our simulations, we solved for the spacecraft's initial state
vectors, unmodeled accelerations, and geophysical parameters of
interest.  We divided our solve-for parameters into two categories:
$local$ and $global$. The $local$ parameters are applicable to a
single flyby only, whereas the $global$ parameters are common to all
flybys.  Each parameter was assigned an a priori uncertainty for the
purpose of our covariance analyses (Table \ref{aprioriPara}).

\begin{table*}
\caption{Solve-for parameters and their a priori uncertainties used in covariance analyses. }
\centering
\begin{tabular*}{\textwidth}{l@{\extracolsep{\fill}} llc}
\hline
Type        &  Parameter                     &  A priori uncertainty      &   Number of parameters   \\ 
\hline
                &    position          &   100 km        & 3 per flyby \\
Local       &     velocity         &   1 m/s        & 3 per flyby \\
                &   constant acceleration &  5$\times$10$^{-11}$ km/s$^2$ & 3 per flyby \\
\hline
                &   GM  & 320 km$^3$/s$^2$ & 1 \\
                &   Love number $k_2$ & 0.3 & 1 \\
 Global     &   20$\times$20 gravity field & Kaula rule (see text) & 437 \\
                &   spin pole  & 1 degree & 2 \\
                &   rotation rate & 1$\times$10$^{-4}$ degree/day & 1 \\
\hline                
Total       & & & 856\\
\hline
\label{aprioriPara}
\end{tabular*}
\end{table*}

We placed a priori constraints on the uncertainties of coefficients of
degree $l>2$ in the expansion of the gravity field.  The constraints
follow a Kaula rule and we adopted the formulation given by
\citet{park2015}:
\begin{equation}
\label{kaula}
K = \frac{(28\times10^{-5})}{l^2}\bigg(\frac{R_m}{R}\bigg)^{l},
\end{equation}
\noindent where $K$ is the a priori constraint for coefficients of
degree $l$ and $R_m$ is the assumed mantle radius of Europa (1465 km).

All $local$ parameters are estimated for each flyby. The constant
acceleration is necessary in order to account for unmodeled
non-gravitational forces (e.g., solar radiation pressure, Jupiter
radiation pressure, etc.).  Its components are expressed in the
Radial-Transverse-Normal (RTN) frame associated with the spacecraft
trajectory.

\subsection{Covariance analysis}
\label{sec-covariance}

To evaluate the precision with which geophysical parameters of Europa
can be determined, we performed covariance analyses \citep{Bierman77}.

Given $z$ observables and $p$ solve-for parameters, the normal
equations can be written as:
\begin{equation}
\label{nomEq}
\eta = H^TWH + {C_{0}}^{-1},
\end{equation}
where $H$ is the matrix of partial derivatives of $z$ with respect to
$p$, $W$ is the matrix of weights appropriate for $z$, and $C_{0}$ is
the a priori covariance matrix of $p$.  We computed and stored the
elements of the normal equations for all flybys, using 11,040 Doppler
measurements, 112 crossover measurements, and 856 solve-for
parameters.  We computed the covariance matrix as:
\begin{equation}
\label{covEq}
 C = \eta^{-1}.
\end{equation}
The formal uncertainties in the estimated parameters are obtained by
taking the square root of the diagonal elements of the covariance
matrix:
\begin{equation}
\label{covEq}
 \sigma_{i} = \sqrt{C_{ii}}.
\end{equation}

The GSWG emphasized that formal uncertainties from covariance analyses
are typically too optimistic, i.e., too small.  The GSWG recommended
multiplying formal uncertainties by a factor of 2 to arrive at more
realistic one-standard-deviation uncertainties.  In this work, we
consistently multiplied formal uncertainties by a factor of 2 per
GSWG recommendations.  All uncertainty values listed or displayed have
the factor of 2 applied.

The covariance analysis technique quickly enables the examination of a
variety of scenarios.  The normal equations are computed and stored
once and for all. If one chooses to restrict the analysis to certain
observables or certain parameters, one simply selects the relevant
subset of lines and columns in $\eta$ (equation~\ref{nomEq}) and
performs a new matrix inversion.

\section{Results}
\label{res}

We assumed that the spacecraft is tracked within $\pm$2 hours of each
closest approach, when the altitude of the spacecraft with respect to
Europa's surface is $\le$ 28,000 km.  The radio link budget depends on
the DSN assets that are deployed to track the spacecraft and on the
spacecraft telecommunication assets, including antenna gain patterns.
The Clipper spacecraft design currently includes two low-gain antennas
and three medium-gain antennas.  We assumed that Doppler measurements
are available only when the radio link budget exceeds a nominal value
(4 dB-Hz) that enables the establishment of a coherent, two-way link.
Our calculations included the spacecraft position and attitude
relevant to the 17F12v2 trajectory, variations due to spacecraft
antenna gain patterns, occultations by Europa and Jupiter, and other
assumptions listed in Table~\ref{tab-assumptions}.

We examined two scenarios. In the first scenario, we assumed that one
of the three most sensitive DSN antennas, the 70 m diameter antennas
at Goldstone, Madrid, and Canberra, was used to track the spacecraft.
This scenario reveals the best performance that one can hope to
achieve with typical ground assets.  In the second scenario, we
assumed that 34 m diameter antennas were deployed either as single
antennas or as arrays of antennas, and we examined the minimum set of
assets that are necessary to meet the gravity science objectives.

We conducted three separate case studies in each of the two scenarios.
In the first case study, we examined the precision that is achieved as
the data from each consecutive flyby is processed.  In the second case
study, we asked how many flybys are necessary to meet the required
measurement precision on $k_2$, $\bar{C}_{20}$, and $\bar{C}_{22}$ if
the flybys tracked with DSN antennas are selected randomly from the
set of all available flybys.  In the third case study, we quantified
the minimum number of flybys that are necessary to meet the
measurement objectives when the flybys are grouped according to their
sub-spacecraft latitude at closest approach.  Results from these case
studies allowed us to gain a progressively deeper understanding of the
measurement precision that can be achieved in a variety of
circumstances.

\subsection{Scenario 1: 70 m DSN antennas}
\label{fulTrk}

\begin{figure}[!htbp]
\centering
\noindent
\includegraphics[width=19pc]{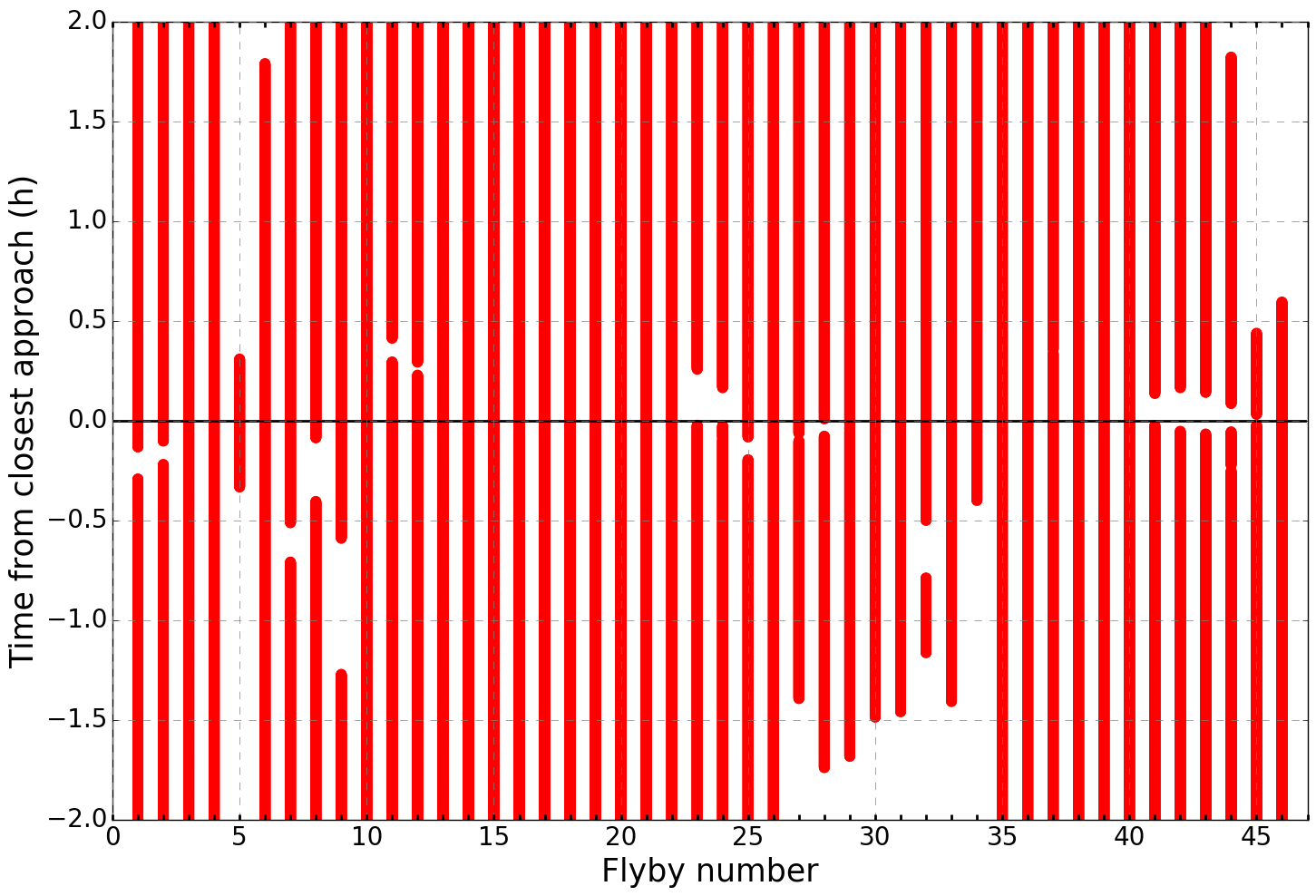}
\caption{Time intervals during which a 4 dB-Hz radio link can be
  maintained between 70 m antennas and Clipper on trajectory 17F12v2.}
\label{rb70m}  
\end{figure}

In Scenario 1, we considered that 70 m DSN antennas were available for
tracking, and we analyzed all 46 flybys of trajectory 17F12v2.  After
applying a 4 dB-Hz cutoff to the radio link budget, we were left with
a total of 10,048 Doppler measurements (Figure \ref{rb70m}).   We also
considered up to 112 crossover measurements (section \ref{croMsr}).
The exact number of Doppler and crossover measurements included in our
analysis depends on the specific flyby selections in the various case
studies.

We note that all flybys in 17F12v2 except E5 can be tracked for at
least 1 h within $\pm$2 h of closest approach with a 70 m antenna.
Flybys with less than 1 h of tracking time are problematic: they
generally contribute little to the realization of measurement
objectives and they have a high ratio of DSN overhead time to useful
tracking time.  Therefore, in this work, we focus on flybys that can
be tracked for a total duration of at least 1 h (not necessarily
continuous).

\subsubsection{Case Study 1: Consecutive flybys}
\label{inv1}

In this case study, we examined the precision of the estimates as
data from consecutive flybys becomes available.  At step $n$, the
available observables consist of observables acquired during flybys 1
through $n$, and the solve-for parameters consist of global parameters
and local parameters relevant to flybys 1 through $n$.

\begin{figure*}
\centering
\noindent
\includegraphics[width=35pc]{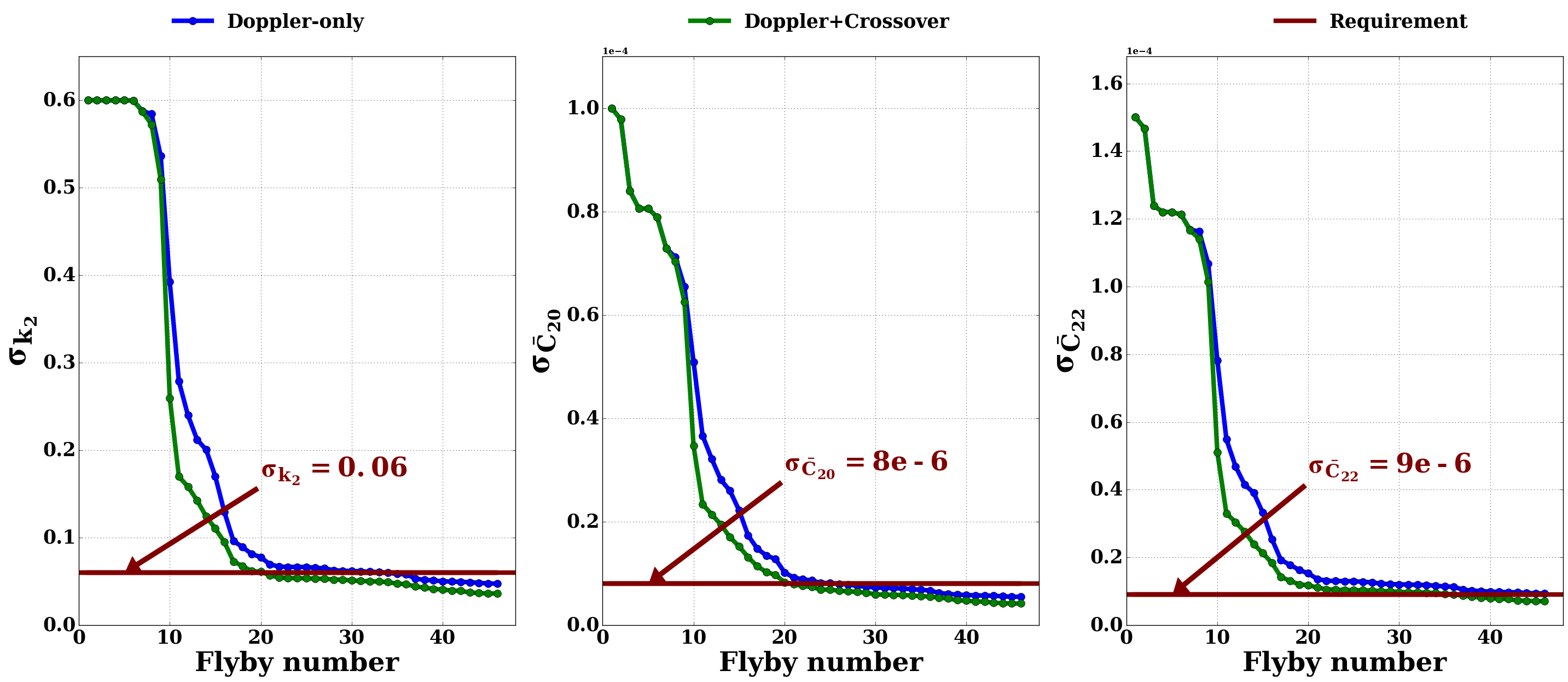}
\caption{Precision of the tidal Love number $k_2$ and gravity field
  coefficients $\bar{C}_{20}$ and $\bar{C}_{22}$ using Doppler (blue)
  and Doppler+crossover (green) measurements when data from 46
  consecutive flybys tracked with 70~m antennas are analyzed (Scenario
  1, Case Study 1). The brown horizontal lines indicate the
  measurement objectives specified in Table \ref{tab-gswg}.  }
\label{seqRec}  
\end{figure*}

We conducted simulations for both Doppler-only and Doppler+crossover
situations (Figure \ref{seqRec}).  We found that measurement
requirements for $k_2$ and $\bar{C}_{20}$ can be met (Table
\ref{seqRecTable}).  The precision of the $\bar{C}_{22}$ gravity
coefficient estimates in Doppler-only simulations does not meet the
measurement objective.  We found that $\bar{C}_{30}$ and
$\bar{C}_{40}$ are never determined with sufficient precision to
estimate the ice shell thickness at the $\pm$20\% level.  The results
indicate that $k_2$ and $\bar{C}_{20}$ measurement objectives can be
met with fewer than 46 flybys, provided that tracking of the
spacecraft is accomplished consistently with 70 m antennas.  The
results also demonstrate that measurement objectives can be achieved
with fewer flybys when crossover measurements are included in the
analysis.

\begin{table*}[h]
  \centering
  \caption{Estimated uncertainties in closest-approach radial
    distances, tidal Love number $k_2$, and low-order gravity field
    coefficients with the 46 flybys of Case Study 1 in Scenario 1
    (tracking Clipper on 17F12v2 trajectory with 70~m antennas).
    Entries in bold indicate that the requirements (rightmost column)
    were not met.  Radial distance uncertainties include an indication of
    the spread (median$\pm$standard deviation) across the 46 flybys. }
\begin{tabular*}{\textwidth}{c@{\extracolsep{\fill}} c c c}    
\hline
Parameter&  Doppler-only    &  Doppler+crossovers & Requirement  \\ 
\hline
Radial distances   &21$\pm$57 m& 19$\pm$53 m & --\\ 
$k_2$      &0.047& 0.036 & $\textless$0.06 \\
$\bar{C}_{20}$ &5.4$\times$10$^{-6}$& 4.1$\times$10$^{-6}$ & $\textless$8.0$\times$10$^{-6}$\\
$\bar{C}_{22}$ &\bf{9.2$\times$10$^{-6}$}& 7.0$\times$10$^{-6}$ & $\textless$9.0$\times$10$^{-6}$\\
$\bar{C}_{30}$ &\bf{7.1$\times$10$^{-7}$}& \bf{6.6$\times$10$^{-7}$} & $\textless$4.0$\times$10$^{-7}$\\
$\bar{C}_{40}$ &\bf{7.3$\times$10$^{-7}$}& \bf{6.8$\times$10$^{-7}$} & $\textless$4.0$\times$10$^{-7}$\\
\hline
\label{seqRecTable}
\end{tabular*}
\end{table*}

\subsubsection{Case Study 2: Random sets of flybys}
\label{inv2}

In this case study, we quantified the minimum number of flybys that
are necessary to meet measurement objectives when flybys tracked with
DSN antennas are selected randomly from the set of available flybys.
In actuality, DSN scheduling would likely take into consideration the
flybys that provide the best possible science return.  For this
reason, we eliminated 4 flybys (E1, E25, E26, E46) that do not
approach Europa's surface within

100 km and 3 flybys (E5, E6, E7) with SEP angles $<$ 20$^\circ$.
These seven flybys are expected to be less valuable than others from a
gravity science perspective. Thus, we considered a maximum of 39
flybys in this case study.  Because the previous case study revealed
the value of combining Doppler and crossover measurements, we included
both types of measurements in this case study.

\begin{figure*}
\centering
\noindent
\includegraphics[width=35pc]{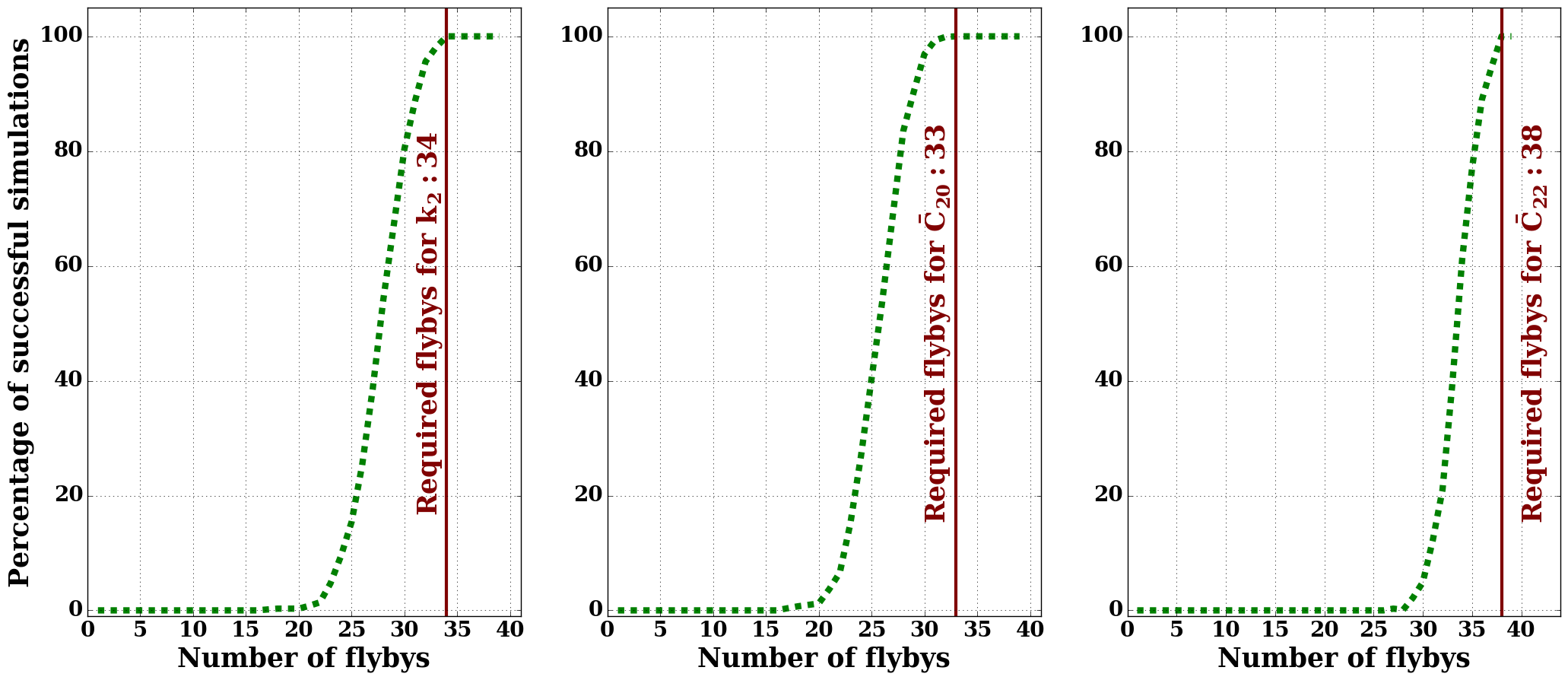}
\caption{Percentage of simulations that meet the tidal Love number
  $k_2$ and gravity field coefficients $\bar{C}_{20}$ and
  $\bar{C}_{22}$ measurement objectives, when considering up to 10,000
  sets of randomly selected 17F12v2 flybys, as a function of the
  number of flybys considered in the sets (Scenario 1, Case Study 2).
  Simulations include both Doppler and crossover measurements and
  incorporate tracking of up to 39 flybys with a 70 m antenna,
  excluding high-altitude
  ($>$ 100  km) and low SEP ($<$ 20$^\circ$) flybys.}
\label{fig-random}  
\end{figure*}

Our simulations used a Monte Carlo scheme in which we considered $n_c$
randomly selected flybys out of $n_a$ available flybys (here,
$n_a=39$).  The number of possible combinations is ($1<= n_c <= 39$):
\begin{equation}
\label{seqPar}
N = \frac{n_a!}{n_c!(n_a-n_c)!}.
\end{equation}
If the number of combinations $N$ was smaller than 10,000, we examined
all possible combinations.  Otherwise, we randomly selected 10,000
cases from the pool of available combinations.  We gradually increased
the number of considered flybys from 1 to 39.
We found that it is possible to meet measurement objectives for $k_2$,
$\bar{C}_{20}$, and $\bar{C}_{22}$ 100\% of the time with 34, 33, and
38 flybys, respectively (Figure \ref{fig-random}).

An important goal of this case study was to gain information about the
distribution of sub-spacecraft latitudes at closest approach that
provides the best prospect of meeting measurement objectives.  Based
on our extensive set of simulations, we were able to quantify the
number of flybys that are required according to specific latitude
regions when considering sets of randomly selected flybys (Table
\ref{tab-random}).

\begin{table*}
\centering
\caption{
  Observed latitudinal distribution of 17F12v2 flybys in successful simulations,
  i.e., in simulations where sets of randomly selected flybys always
  met the tidal Love number $k_2$ and $\bar{C}_{20}$ and
  $\bar{C}_{22}$ gravity field coefficients measurement
  objectives (Scenario 1, Case Study 2).  Simulations include both
  Doppler and crossover measurements and incorporate tracking of up to
  39 flybys with a 70 m antenna, excluding high-altitude
  ($>$ 100 km) and low SEP ($<$
  20$^\circ$) flybys.  The rightmost columns indicate the medians and
  standard deviations of the number of flybys that were included in
  successful simulations.} 
\begin{tabular*}{\textwidth}{l@{\extracolsep{\fill}} c c c c c }
\hline
Europa region &  Latitude range & Available flybys & $k_2$    &  $\bar{C}_{20}$ & $\bar{C}_{22}$  \\  \hline
High latitude north  &    90$^\circ$ $\textendash$  45$^\circ$  & 8   &7$\pm$1 & 7$\pm$1 & 7$\pm$0\\
Mid latitude north    &   45$^\circ$ $\textendash$  15$^\circ$  & 4   &4$\pm$1 & 3$\pm$1 & 4$\pm$0\\
Low latitude          &   15$^\circ$ $\textendash$ -15$^\circ$  & 13   &11$\pm$1 & 11$\pm$1 & 13$\pm$1\\
Mid latitude south    &  -15$^\circ$ $\textendash$ -45$^\circ$  & 8  &7$\pm$1 & 7$\pm$1 & 8$\pm$0\\
High latitude south  &   -45$^\circ$ $\textendash$ -90$^\circ$  & 6  &5$\pm$1 & 5$\pm$1 & 6$\pm$0\\
\hline
Total                & 90$^\circ$ $\textendash$ -90$^\circ$& 39 & 34 & 33 & 38\\
\hline
\label{tab-random}
\end{tabular*}
\end{table*}

\subsubsection{Case Study 3: Preferred sets of flybys}
\label{inv3}

In this third and final case study of Scenario 1, we used knowledge
gained in previous case studies to inject some intelligence in the
selection of flybys that can meet the primary ($k_2$) measurement
objective.  As in the previous case study, we discarded 7 flybys that
either have low ($<$ 20$^{\circ}$) SEP angles or high ($>$ 100 km)
closest-approach altitudes.  The remaining 39 flybys were classified
into latitude regions.  Because of the arrangement of the fan-beam
(medium gain) antennas on the spacecraft, the low-latitude flybys are
easiest to track with DSN assets.  Thus, our selection started with
all 13 low-latitude flybys.  In the first step, we evaluated the
performance with all 13 low-latitude flybys plus a randomly selected
flyby from each mid-latitude band, for a total of 15 flybys.  In the
next step, we considered all 13 low-latitude flybys, a randomly
selected flyby from each mid-latitude band, and a randomly selected
flyby from each high-latitude band, for a total of 17 flybys.  We
gradually increased the number of mid- and high-latitude flybys in
this fashion.  At each step, we examined all possible combinations of
flybys.  We found that, with 70 m DSN assets, it is possible to meet
the $k_2$ measurement objective with 23 flybys (Table
\ref{tab-preferred}).
Referring back to Figure~\ref{fig-random}, about 10\% of
the simulations with 23 randomly selected flybys were successful with
respect to $k_2$.

\begin{table*}
\centering
\caption{Number of methodically selected 17F12v2 mid- and
  high-latitude flybys required to meet the tidal Love number $k_2$
  measurement objective when flybys are tracked with 70~m DSN antennas
  and all low-latitude flybys are tracked (Scenario 1, Case Study 3).}
  \begin{tabular*}{\textwidth}{l@{\extracolsep{\fill}}c c c}
\hline
Europa region &  Latitude range & Avail. flybys & Req. flybys      \\ 
\hline
High latitude north  &    90$^\circ$ $\textendash$  45$^\circ$  & 8&  2\\
Mid latitude north    &   45$^\circ$ $\textendash$  15$^\circ$  & 4& 3\\
Low latitude          &   15$^\circ$ $\textendash$ -15$^\circ$  & 13 &13\\
Mid latitude south    &  -15$^\circ$ $\textendash$ -45$^\circ$  & 8 & 3\\
High latitude south  &   -45$^\circ$ $\textendash$ -90$^\circ$  & 6 & 2\\
\hline
Total                & 90$^\circ$ $\textendash$ -90$^\circ$& 39 & 23\\
\hline
\label{tab-preferred}
\end{tabular*}
\end{table*}

\begin{figure*}
\centering
\noindent
\includegraphics[width=35pc]{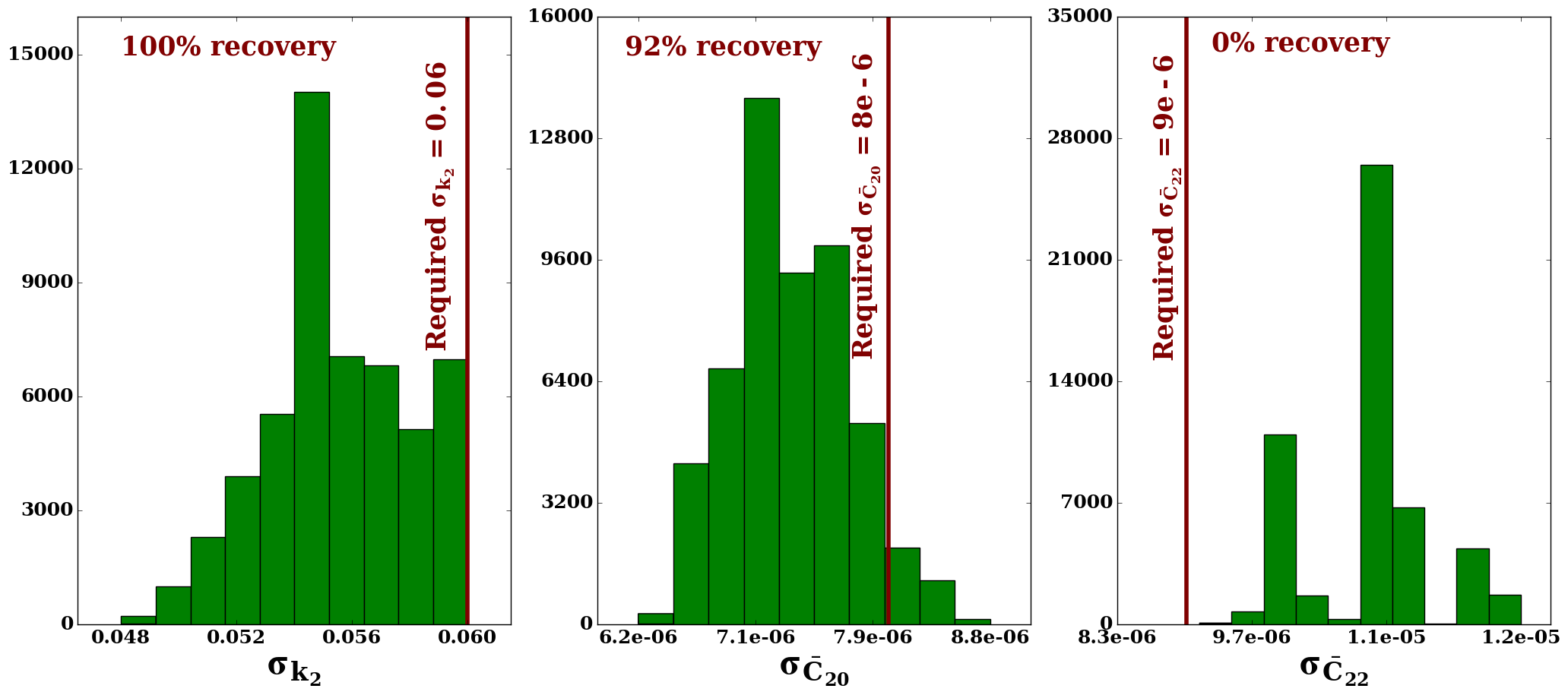}
\caption{Histograms of $k_2$, $\bar{C}_{20}$, and $\bar{C}_{22}$
  uncertainties obtained by performing covariance analyses for all
  possible combinations of 23 flybys with the latitudinal distribution
  shown in Table \ref{tab-preferred} and tracking with 70~m antennas
  over time intervals shown in Figure \ref{rb70m} (Scenario 1, Case
  Study 3).}
\label{preferredRun}  
\end{figure*}

We evaluated $k_2$, $\bar{C}_{20}$, and $\bar{C}_{22}$ parameter
uncertainties with the minimum number of flybys specified in Table
\ref{tab-preferred}.  We performed covariance analyses for all
possible combinations and found that it is possible to meet the $k_2$
measurement objective 100\% of the time, whereas the requirements for
$\bar{C}_{20}$ and $\bar{C}_{22}$ are met only 92\% and 0\% of the
time, respectively.  It is possible to meet the $\bar{C}_{20}$ and
$\bar{C}_{22}$ measurement objectives by increasing the number of
flybys that are tracked (section \ref{inv2}).

\subsection{Scenario 2: Minimum DSN assets}
\label{parTrk}

Scenario 1 provided estimates of what can be achieved with 70 m
antennas.  However, 70 m antenna time is difficult to schedule.  In
Scenario 2, we considered situations that place fewer demands on the
ground telecommunication assets.  We identified the minimal set of
ground assets that achieve the $k_2$ measurement objective.

\begin{figure*}
\centering
\noindent
\includegraphics[width=18pc]{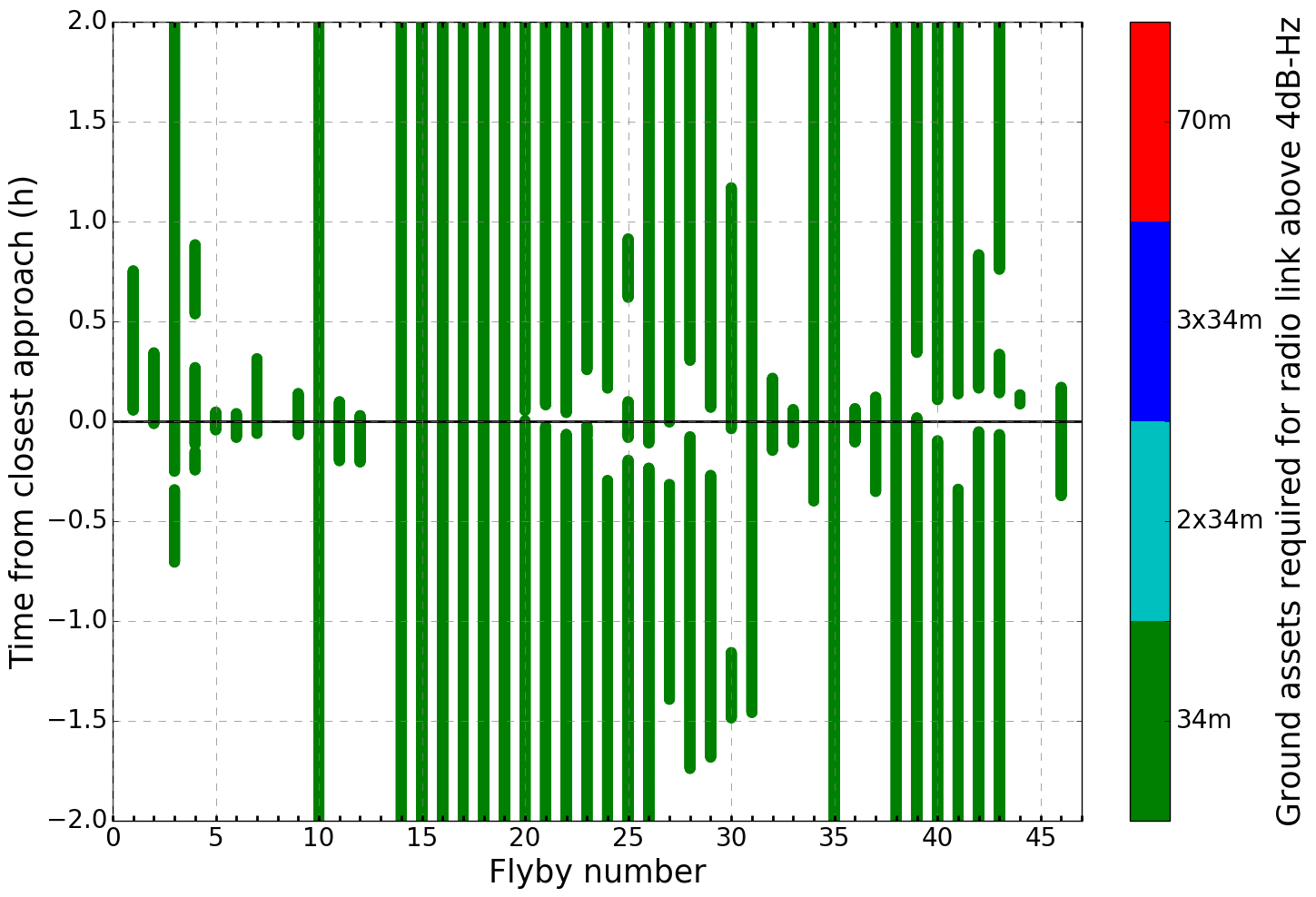}{a}
\includegraphics[width=18pc]{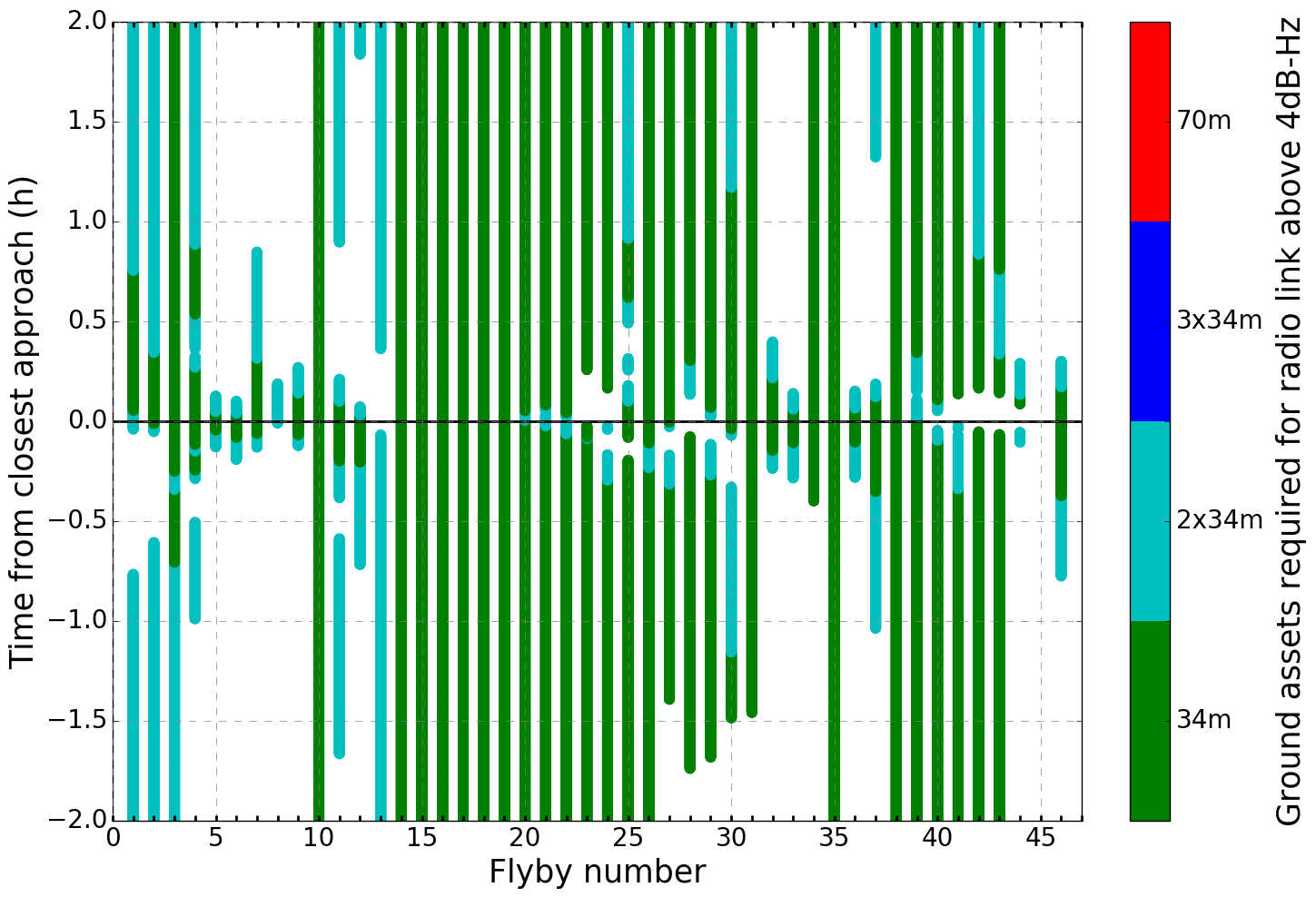}{b}
\includegraphics[width=18pc]{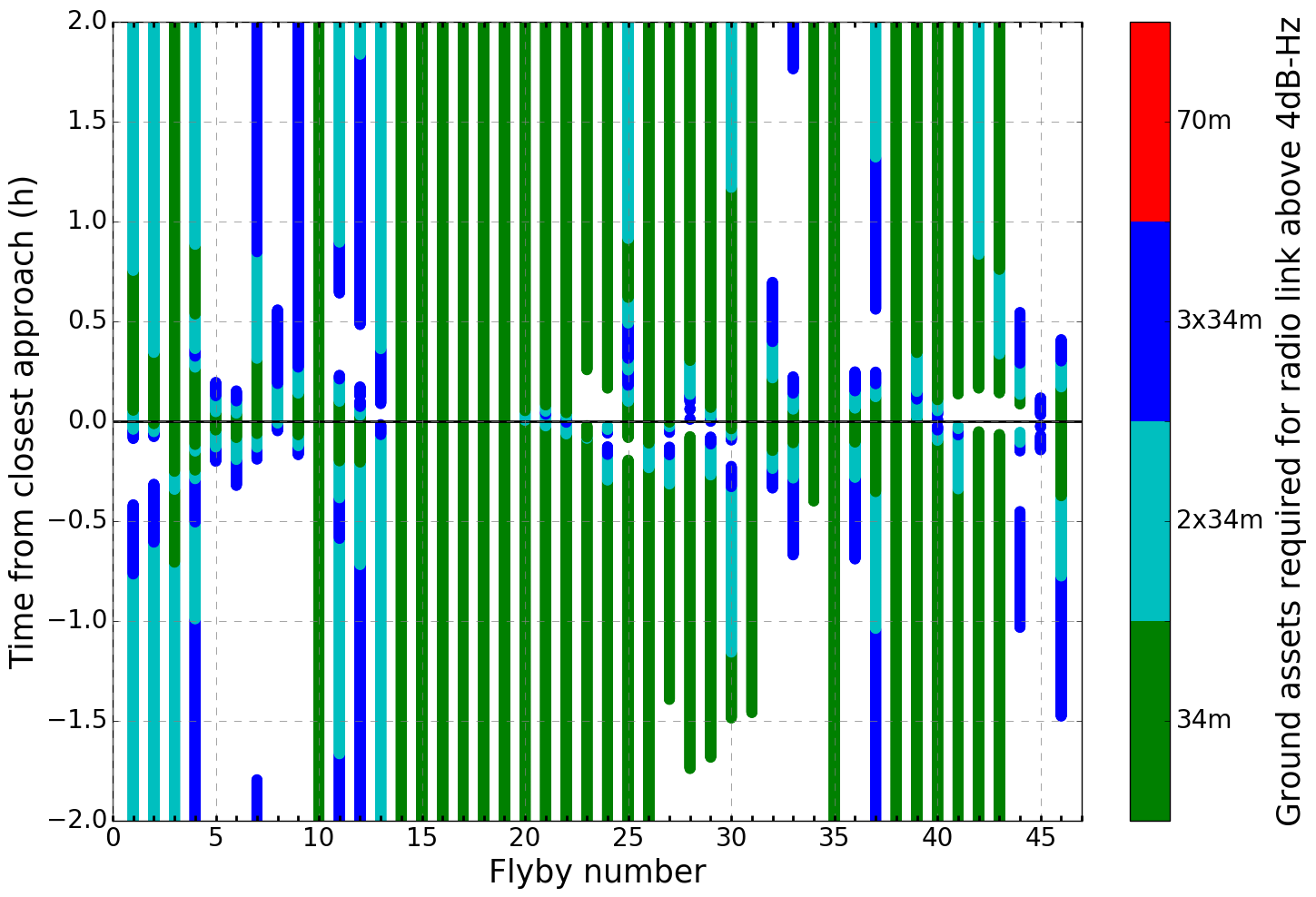}{c}
\includegraphics[width=18pc]{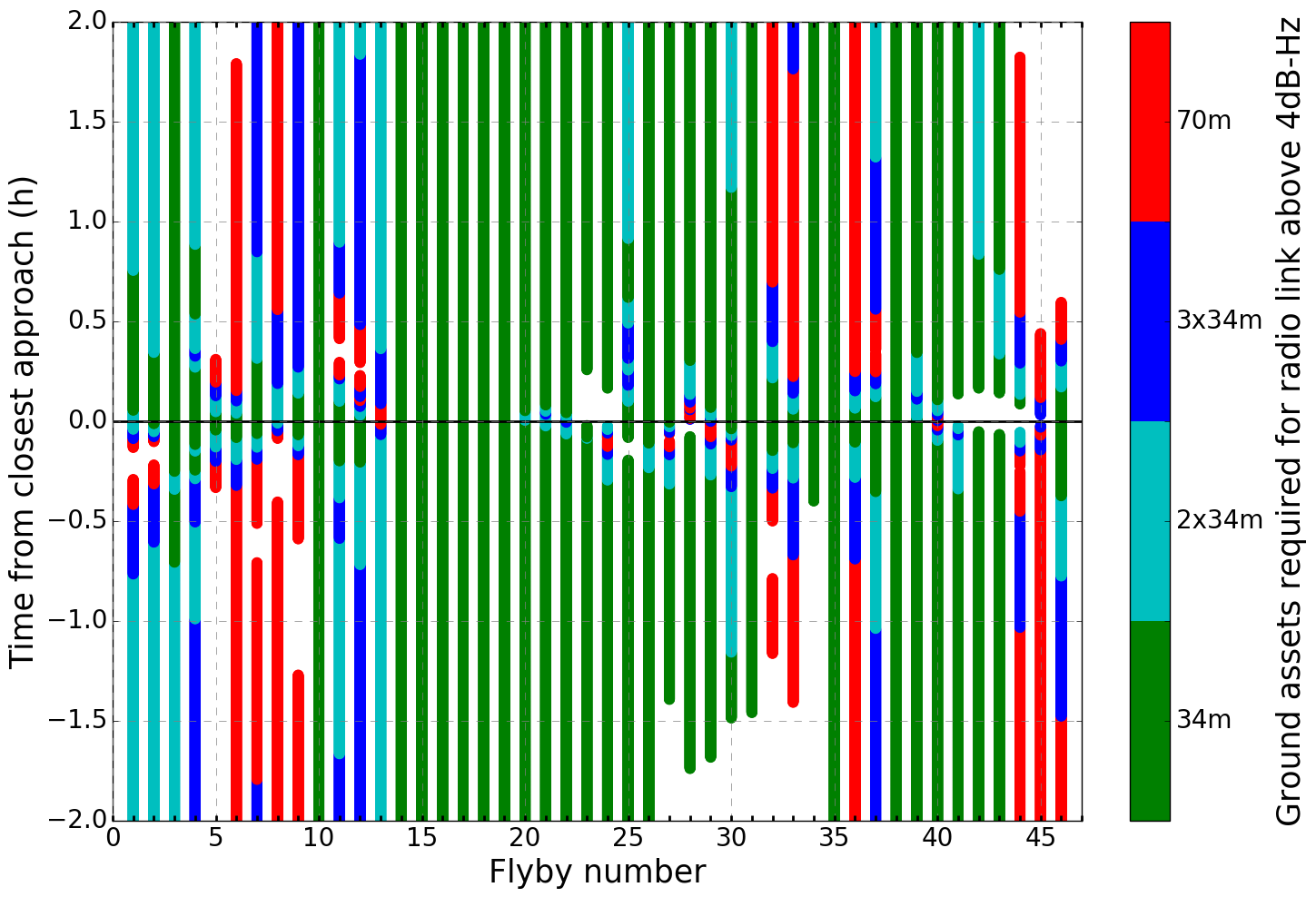}{d}
\caption{Time intervals during which a 4 dB-Hz radio link can be
  maintained between Clipper and progressively more sensitive DSN
  assets: (a) a 34 m antenna, (b) two 34 m antennas, (c) three 34
  m antennas, and (d) a 70 m antenna.}
\label{antGain}  
\end{figure*}

Similar to Scenario 1, we assumed that Doppler measurements were
available only when the radio link budget exceeded 4 dB-Hz within
$\pm$2 hours of each closest approach. One can deploy a variety of
ground assets to maintain such a radio link.  We considered four
configurations: a 34 m antenna, an array composed of two 34 m
antennas, an array composed of three 34 m antennas, and a 70 m
antenna.  For each of these configurations, we computed the time
intervals during which a 4 dB-Hz radio link can be maintained (Figure
\ref{antGain}).

We examined the distribution of flybys according to tracking duration
with various DSN assets (Table \ref{dsnAssetsNotused}).  A 34 m
antenna can track 9 flybys for $\pm$2 hours of each closest approach.
With a two-antenna or three-antenna array, the number of flybys that
can be tracked increases to 12 and 16, respectively.  A 70 m antenna
can track 19 flybys for $\pm$2 hours of each closest approach.

\begin{table}[h]
\centering
\caption{Number of 17F12v2 flybys according to total (not necessarily
  continuous) tracking duration with a 4~dB-Hz link budget and various DSN assets.}
\begin{tabular}{cccc}
\hline
\multirow{2}{*}{DSN assets} & \multicolumn{3}{c}{Number of flybys} \\ \cline{2-4} 
                            & $>$ 4 h   &  1--4 h  & $<$ 1 h\\ \hline
34 m                        & 9         & 19         & 18\\ 
2x34 m                      & 12        & 23         & 11\\ 
3x34 m                      & 16        & 25         & 5 \\ 
70 m                        & 19        & 26         & 1 \\ \hline
\label{dsnAssetsNotused}
\end{tabular}
\end{table}

\begin{figure}[!htbp]
\centering
\noindent
\includegraphics[width=19pc]{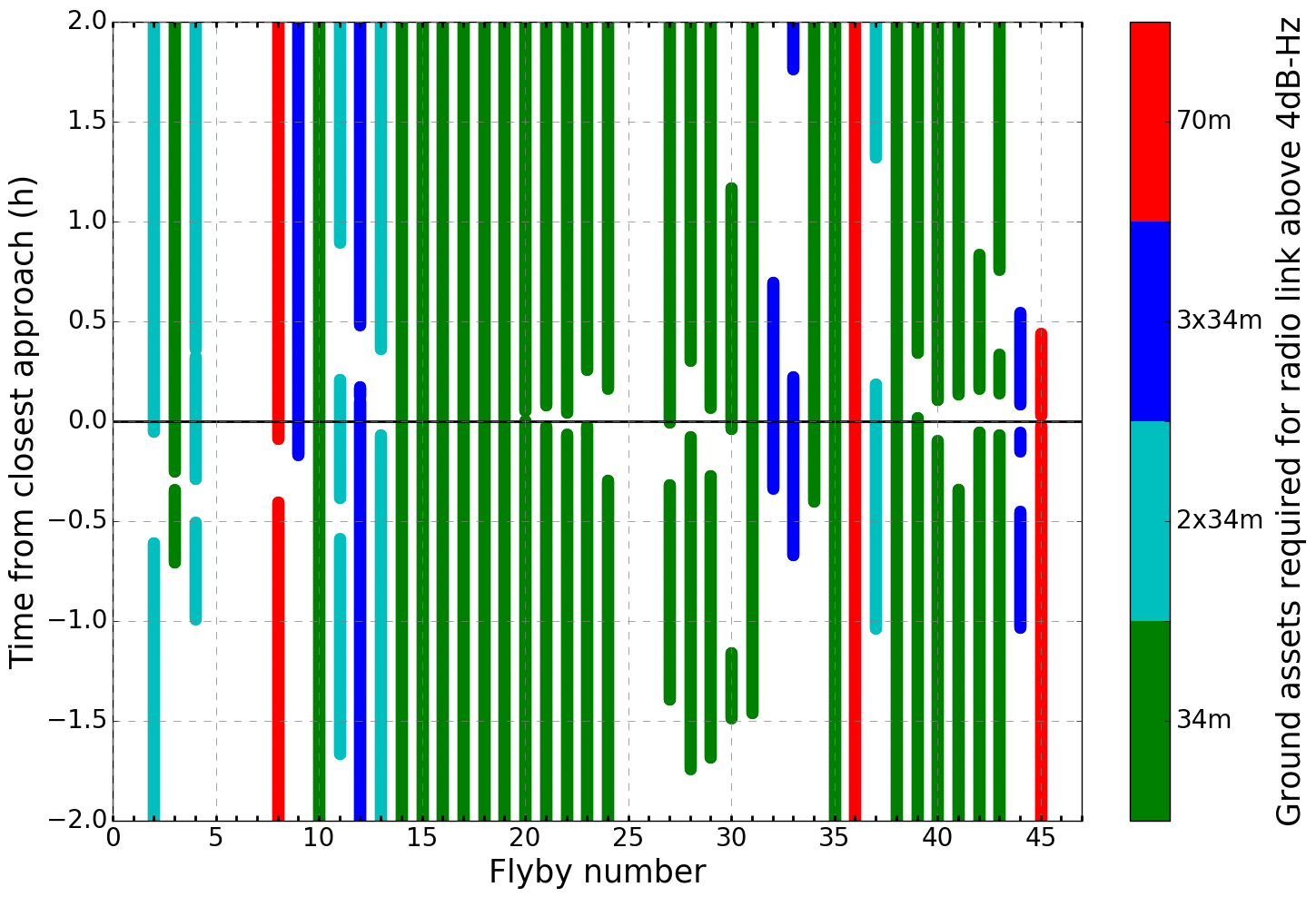}
\caption{ Time intervals during which a 4 dB-Hz radio link can be
  maintained for at least 1 h.  For each flyby, the least sensitive
  antenna configuration was used.  Flybys with low SEP angles
  ($<$20$^\circ$) or high altitudes ($>$100 km) were discarded, leaving
  a total of 39 flybys.  See also Table~\ref{dsnAssets}.}
\label{fig-dsnAssets}  
\end{figure}

As in Scenario 1 (section \ref{fulTrk}), we only considered flybys
that provide the best possible science return. We therefore discarded
7 flybys either with high closest approach altitudes ($>$ 100 km) or
low SEP angles ($<$ 20$^\circ$). For the remaining 39 flybys, we
identified the number of flybys for which Clipper can be tracked for
at least 1 h (not necessarily continuous) within $\pm$2~h of closest
approach with a link budget above 4 dB-Hz.  We discarded flybys with
less than 1 h of total DSN tracking because these flybys generally
contribute little to the realization of measurement objectives and
because of their high ratio of DSN overhead time to useful tracking
time.  A 34~m antenna is sufficient to track 26 out of the 39
considered flybys for more than 1 h.  We label this configuration
DSN$_\textrm{34m}$.  If we combine two 34 m antennas into a
two-antenna array, 5 additional flybys can be tracked for more than 1
h.  The union of DSN$_\textrm{34m}$ and these 5 additional flybys
yields a total of 31 flybys, a configuration that we label
DSN$_\textrm{2$\times$34m}$.  If we combine three 34 m antennas into a
three-antenna array, 5 additional flybys can be tracked for more than
1 h.  The union of DSN$_\textrm{2$\times$34m}$ and these 5 additional
flybys yields a total of 36 flybys, a configuration that we label
DSN$_\textrm{3$\times$34m}$.  If a 70 m antenna is used, 3 additional
flybys can be tracked for more than 1 h.  The union of
DSN$_\textrm{3$\times$34m}$ and these 3 additional flybys yields a
total of 39 tracked flybys, a configuration that we label
DSN$_\textrm{70m}$.  The number of available flybys with each DSN
configuration are summarized in Table \ref{dsnAssets}, and the
corresponding tracking coverage is illustrated in Figure
\ref{fig-dsnAssets}. In order to minimize the use of ground assets, we
select, for each flyby, the least sensitive antenna configuration that
can provide at least 1 h of tracking.  Specifically, we do not use
additional assets to extend the duration of flybys that are already
tracked for at least 1 h (compare Figure~\ref{antGain}(d) to Figure
\ref{fig-dsnAssets}).

\begin{table*}[h]
\centering
\caption{ Number of 17F12v2 flybys that can be tracked ($\geq$
  4~dB-Hz) for at least 1 h (not necessarily continuous, Figure
  \ref{fig-dsnAssets}) within $\pm$2 h of each closest approach for
  increasingly sensitive antenna configurations.}
\begin{tabular*}{\textwidth}{l@{\extracolsep{\fill}}c c c c c}
\hline
\multirow{2}{*}{Configuration} & \multicolumn{4}{c}{Number of available flybys} & \multirow{2}{*}{Total} \\ \cline{2-5}
                           & 34 m         & 2x34 m         & 3x34 m       & 70 m       &\\  \hline
DSN$_\textrm{34m}$                      & 26          & ...           & ...         & ...        & 26                     \\
DSN$_\textrm{2$\times$34m}$         & 26          & 5             & ...         & ...        & 31                     \\
DSN$_\textrm{3$\times$34m}$      & 26          & 5             & 5           & ...        & 36                     \\
DSN$_\textrm{70m}$       & 26          & 5             & 5           & 3          & 39              \\ \hline      
\label{dsnAssets}
\end{tabular*}
\end{table*}

The available number of Doppler and crossover measurements varies
according to the chosen DSN configuration (Table~\ref{msrPoints}).
However, the exact number of Doppler and crossover measurements
included in our analysis depends on the specific flyby selections in
the various case studies.

\begin{table*}[h]
\centering
  \caption{ Number of Doppler and crossover measurements that can be
  obtained with the antenna configurations listed in Table \ref{dsnAssets}, 
  compared to the numbers obtained when tracking 39 flybys with 
  70 m antennas. The number of crossovers corresponds to the number 
  of intersections of tracked flybys. 
  Also shown are the total durations
  during which tracking can be conducted with a link budget above 4
  dB-Hz, expressed as a fraction of the total potential tracking time
  ($\pm$2 h of each closest approach, or 184 h in 17F12v2). Flybys
  with $<$ 1 h tracking duration, SEP angle $<$ 20$^\circ$, and
  altitude $>$ 100 km were discarded.}
  \begin{tabular*}{\textwidth}{l@{\extracolsep{\fill}}c c c}
  \hline 
DSN               & Doppler            & Crossover        & Tracking  \\
config.           & msr.               & msr.             & fraction  \\ \hline
DSN$_\textrm{34m}$                      & 5503  & 37      & 50\%            \\ 
DSN$_\textrm{2$\times$34m}$      & 6371  & 62      & 58\%             \\ 
DSN$_\textrm{3$\times$34m}$      & 6922   & 84      & 63\%             \\ 
DSN$_\textrm{70m}$                     & 7527   & 95      & 68\%         \\  
\hline
Tracking of 39 flybys with 70 m antennas  & 8756 & 95 & 79\%\\
\hline
\label{msrPoints}
\end{tabular*}
\end{table*}

\subsubsection{Case Study 1: Consecutive flybys}
\label{inv1-2x34}

This Scenario 2 case study is based on the same principles as its
analog in Scenario 1 (section \ref{inv1}). We analyzed the Doppler and
crossover data from consecutive flybys as it becomes available.

\begin{figure*}
\centering
\noindent
\includegraphics[width=35pc]{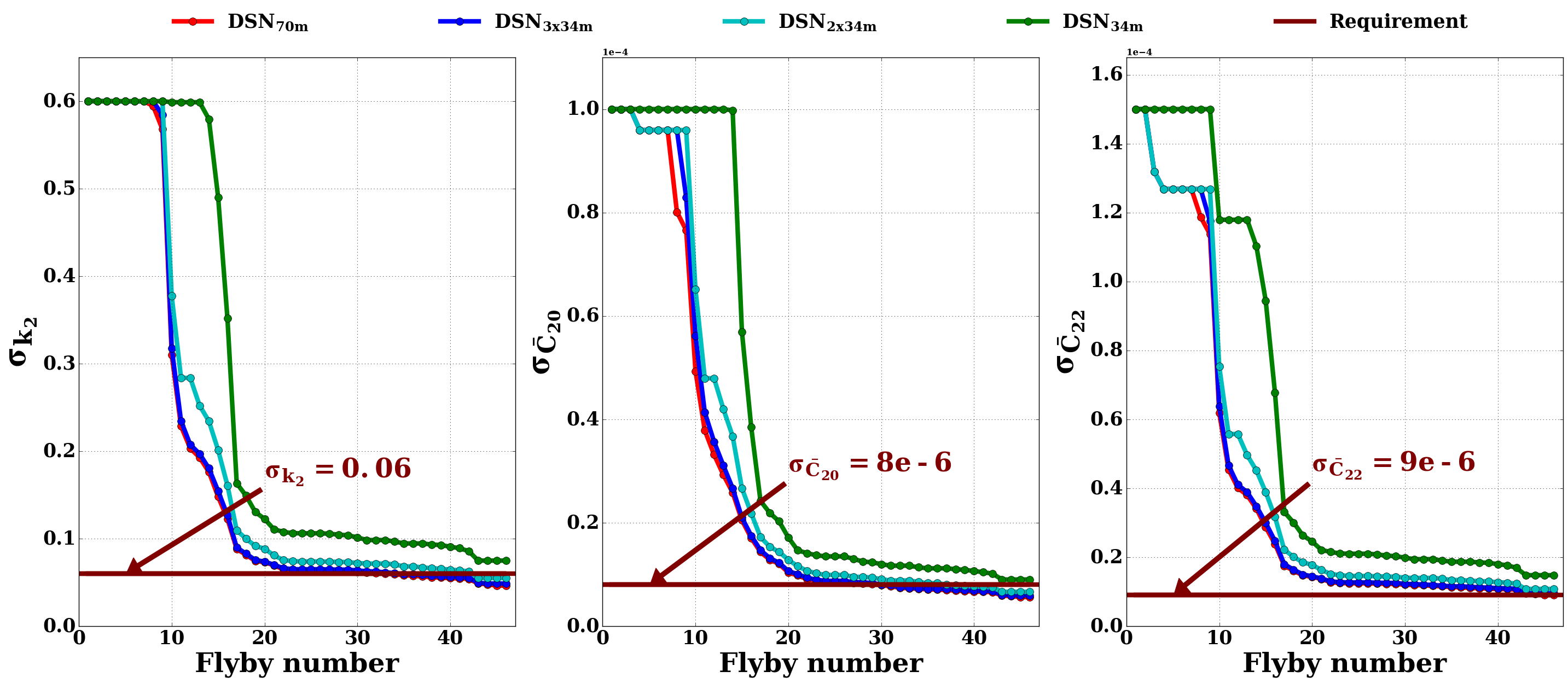}
\caption{Precision of the tidal Love number $k_2$ and gravity field
  coefficients $\bar{C}_{20}$ and $\bar{C}_{22}$ estimates when data from 46
  consecutive flybys are analyzed (flybys with $<$ 1 h tracking
  duration, SEP angle $<$ 20$^\circ$, and altitude $>$ 100 km were
  discarded) using progressively more sensitive DSN configurations
  (Scenario 2, Case Study 1, Table \ref{dsnAssets}).  The green curve
  shows the performance with 26 flybys tracked with a single 34~m
  antenna configuration (DSN$_\textrm{34m}$).  The cyan curve
  considers the addition of a second antenna for 5 flybys, for a total
  of 31 flybys (DSN$_\textrm{2$\times$34m}$).  The blue curve
  considers the addition of a third antenna for 5 flybys, for a total
  of 36 flybys (DSN$_\textrm{3$\times$34m}$).  The red curve considers
  the addition of a 70~m antenna for 3 flybys, for a total of 39
  flybys (DSN$_\textrm{70m}$). The red curve is nearly
  indistinguishable from the blue curve.  The brown horizontal lines
  indicate the measurement objectives specified in Table
  \ref{tab-gswg}.  }
\label{seqRec2x34m}  
\end{figure*}

We found that measurement objectives cannot be met if tracking is
restricted to a single 34 m antenna (Figure \ref{seqRec2x34m}).
However, it is possible to meet the $k_2$ measurement objective with
the DSN$_\textrm{2$\times$34m}$ configuration by tracking 26 flybys
with a single 34 m antenna and 5 additional flybys with a two-antenna
array (2x34 m), for a total of 31 tracked flybys.

The estimated uncertainties in tidal Love number $k_2$ and low-order
gravity field coefficients using a variety of DSN assets are shown in
Table \ref{seqRecTable2x34m}. The measurement objective pertaining to
verifying whether the ice shell is hydrostatic (Table \ref{tab-gswg})
is barely met with the most sensitive antenna configuration.  The
gravity field coefficients $\bar{C}_{30}$ and $\bar{C}_{40}$ are never
measured at the level required to measure the ice shell thickness with
$\pm$20\% uncertainties (section~\ref{intro}).

\begin{table*}
\centering
\caption{Estimated uncertainties in tidal Love number $k_2$ and
  low-order gravity field coefficients with the tracked ($>$ 1 h
  duration) 17F12v2 flybys of Case Study 1 in Scenario 2, as a
  function of DSN configuration (Table~\ref{dsnAssets}).  Entries in
  bold indicate that the requirement (rightmost column) was not met.}
\begin{tabular}{c c c c c c}
\hline
\multirow{2}{*}{Parameter} &   DSN$_\textrm{34m}$    & DSN$_\textrm{2$\times$34m}$  & DSN$_\textrm{3$\times$34m}$  &  DSN$_\textrm{70m}$ & \multirow{2}{*}{Requirement} \\
             & (26 flybys) & (26+5=31 flybys) & (26+5+5=36 flybys) & (26+5+5+3=39 flybys)\\
\hline
$k_2$         &\bf{0.075}                & 0.055     &0.049 & 0.046 & $\textless$0.06 \\
$\bar{C}_{20}$ &\bf{9.0$\times$10$^{-6}$} & 6.6$\times$10$^{-6}$  &6.0$\times$10$^{-6}$ &5.6$\times$10$^{-6}$ &$\textless$8.0$\times$10$^{-6}$\\
$\bar{C}_{22}$ &\bf{15$\times$10$^{-6}$}& \bf{11$\times$10$^{-6}$} & \bf{9.5$\times$10$^{-6}$} &\bf{9.0$\times$10$^{-6}$} &$\textless$9.0$\times$10$^{-6}$\\
$\bar{C}_{30}$ &\bf{33$\times$10$^{-7}$}& \bf{26$\times$10$^{-7}$}  &  \bf{21$\times$10$^{-7}$} & \bf{18$\times$10$^{-7}$} &$\textless$4.0$\times$10$^{-7}$\\
$\bar{C}_{40}$ &\bf{31$\times$10$^{-7}$}& \bf{27$\times$10$^{-7}$}  &  \bf{20$\times$10$^{-7}$} & \bf{16$\times$10$^{-7}$} &$\textless$4.0$\times$10$^{-7}$\\
\hline
\label{seqRecTable2x34m}
\end{tabular}
\end{table*}

\subsubsection{Case Study 2: Random sets of flybys}
\label{inv2-2x34}

This Scenario 2 case study is based on the same principles as its
analog in Scenario 1 (section \ref{inv2}).  We quantified the minimum
number of flybys that are necessary to meet measurement objectives
when tracked ($>$ 1 h) flybys are selected randomly from the set of available flybys.  

Because it is not possible to meet science objectives with a single 34
m antenna (DSN$_\textrm{34m}$), we selected the
DSN$_\textrm{2$\times$34m}$ antenna configuration, where a 2x34 m
antenna array is used on up to 5 occasions to supplement the up to 26
flybys tracked with a single 34 m antenna.  Thus, a total of up to 31
tracked flybys are available in this case study (Table
\ref{dsnAssets}).

\begin{figure*}
\centering
\noindent
\includegraphics[width=35pc]{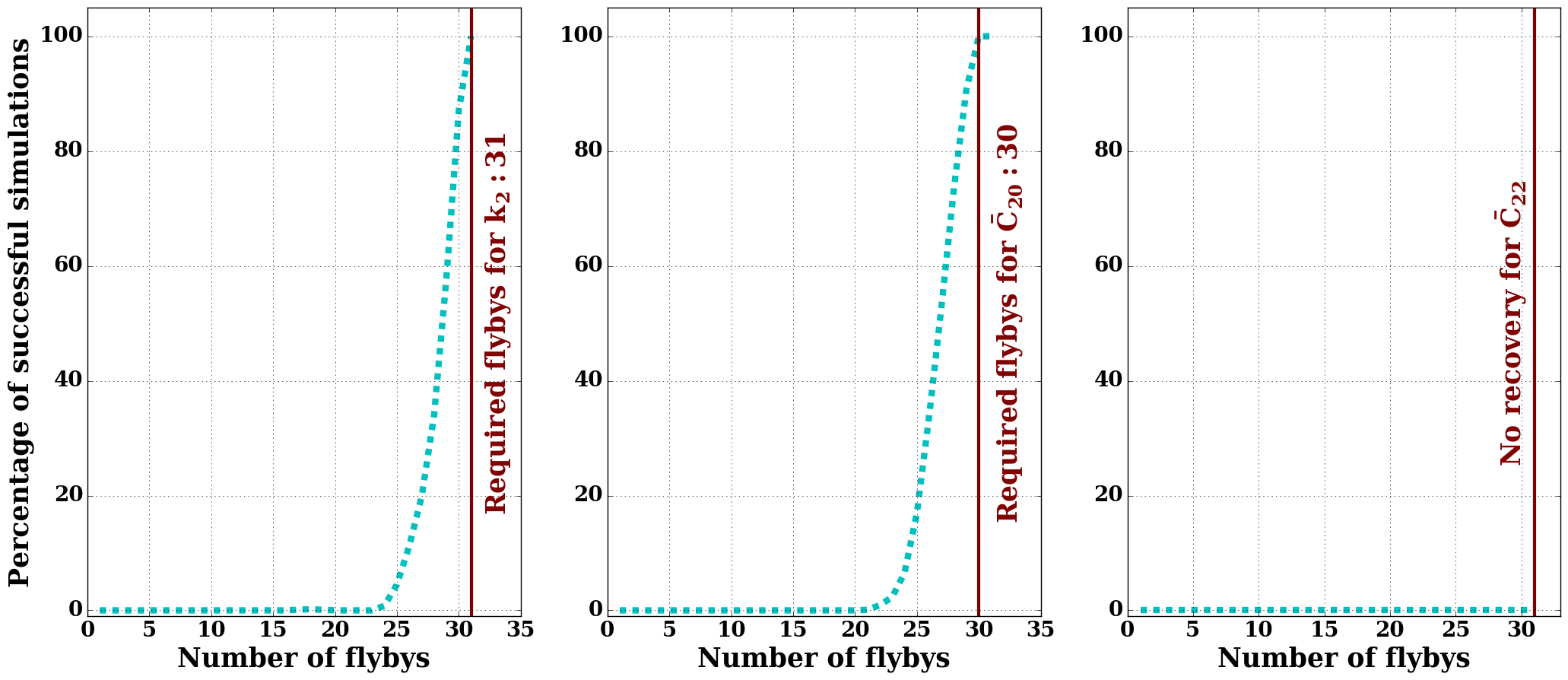}

\caption{Percentage of simulations that meet the tidal Love number
  $k_2$ and gravity field coefficients $\bar{C}_{20}$ and
  $\bar{C}_{22}$ measurement objectives, when considering up to 10,000
  sets of randomly selected, tracked ($>$ 1 h duration), 17F12v2
  flybys, as a function of the number of flybys considered in the sets
  (Scenario 2, Case Study 2).  Simulations include both Doppler and
  crossover measurements and incorporate tracking of up to 31 flybys
  with the DSN$_\textrm{2$\times$34m}$ antenna configuration (26
  flybys tracked with a single 34 m antenna and 5 additional flybys
  tracked with a 2x34 m antenna array), excluding high-altitude
  ($>$ 100 km) and low SEP ($<$ 20$^\circ$) flybys.}
\label{fig-random2x34m}  
\end{figure*}

We considered $n_c$ randomly selected flybys out of $n_a$ available
flybys (here, $n_a=31$).  The number of possible combinations is given
by equation \ref{seqPar}.  If the number of combinations $N$ was
smaller than 10,000, we examined all possible combinations.  Otherwise,
we randomly selected 10,000 cases from the pool of available
combinations.  We gradually increased the number of considered flybys
from 1 to 31.

We found that it is possible to meet the $k_2$ and $\bar{C}_{20}$
measurement objectives 100\% of the time with 31 and 30 randomly
chosen flybys, respectively.  The measurement objective for
$\bar{C}_{22}$ was never achieved (Figure~\ref{fig-random2x34m}).  The
counter-intuitive result of meeting the $k_2$ ($\bar{C}_{20}$)
measurement objectives with 31 (30) flybys in this scenario versus 34
(33) flybys in the Case Study 2 of Scenario 1 (section \ref{inv2}) is
due to the fact that, in Scenario 2, a larger proportion of equatorial
flybys is represented in the pool of 31 flybys than in the pool of 39
flybys in Scenario 1.  Equatorial flybys provide a better
determination of $k_2$ than high-latitude flybys.

As in Scenario 1, we identified the latitudinal distribution of flybys
in successful simulations.  (Table \ref{tab-random2x34m}).

\begin{table*}[!htbp]
\centering
\caption{
  Observed latitudinal distribution of 17F12v2 flybys in
  successful simulations, i.e., in simulations where sets of randomly
  selected flybys always met the tidal Love number $k_2$ and
  $\bar{C}_{20}$ gravity field coefficients
  measurement objectives (Scenario 2, Case Study 2).    The measurement objective for
  $\bar{C}_{22}$ was never achieved.
  Simulations include both Doppler and crossover measurements and
  incorporate tracking of up to 31 flybys with the
  DSN$_\textrm{2$\times$34m}$ antenna configuration (26 flybys tracked
  with a single 34~m antenna and 5 additional flybys tracked with a
  two-antenna array), excluding high-altitude
  ($>$ 100 km) and low SEP ($<$ 20$^\circ$) flybys.
  The rightmost columns indicate the medians and standard deviations of
  the number of flybys that were included in successful simulations.
}
\begin{tabular*}{\textwidth}{l@{\extracolsep{\fill}} c c c c c}
\hline
Europa region &  Latitude range &Available flybys & $k_2$    &  $\bar{C}_{20}$ & $\bar{C}_{22}$  \\ 
\hline
High latitude north  &    90$^\circ$ $\textendash$  45$^\circ$  & 7   &7$\pm$0 & 6$\pm$1 & $...$\\
Mid latitude north    &   45$^\circ$ $\textendash$  15$^\circ$  & 4   &4$\pm$0 & 4$\pm$0 & $...$\\
Low latitude          &   15$^\circ$ $\textendash$ -15$^\circ$  & 13   &13$\pm$0 & 13$\pm$0 & $...$\\
Mid latitude south    &  -15$^\circ$ $\textendash$ -45$^\circ$  & 5  &5$\pm$0 & 5$\pm$0 & $...$\\
High latitude south  &   -45$^\circ$ $\textendash$ -90$^\circ$  & 2  &2$\pm$0 & 2$\pm$0 & $...$\\
\hline
Total                & 90$^\circ$ $\textendash$ -90$^\circ$& 31 & 31 & 30 & $...$\\
\hline
\label{tab-random2x34m}
\end{tabular*}
\end{table*}

\subsubsection{Case Study 3: Preferred sets of flybys}
\label{inv3-2x34}

In this third and final case study for Scenario 2, we examined the
number of tracked ($>$ 1 h duration) flybys required to meet the $k_2$
measurement objective with careful selection of the flybys according
to latitude region. As shown in section \ref{inv1-2x34}, a single 34~m
antenna is not sufficient to meet this objective. Therefore we
selected the DSN$_\textrm{2$\times$34m}$ antenna configuration (Table
\ref{dsnAssets}) to perform this case study with 31 tracked flybys.

As in section~\ref{inv3}, we started with all low-latitude flybys and gradually
increased the number of randomly selected mid- and high-latitude
flybys until the measurement objective was achieved.

\begin{figure*}
\centering
\noindent
\includegraphics[width=35pc]{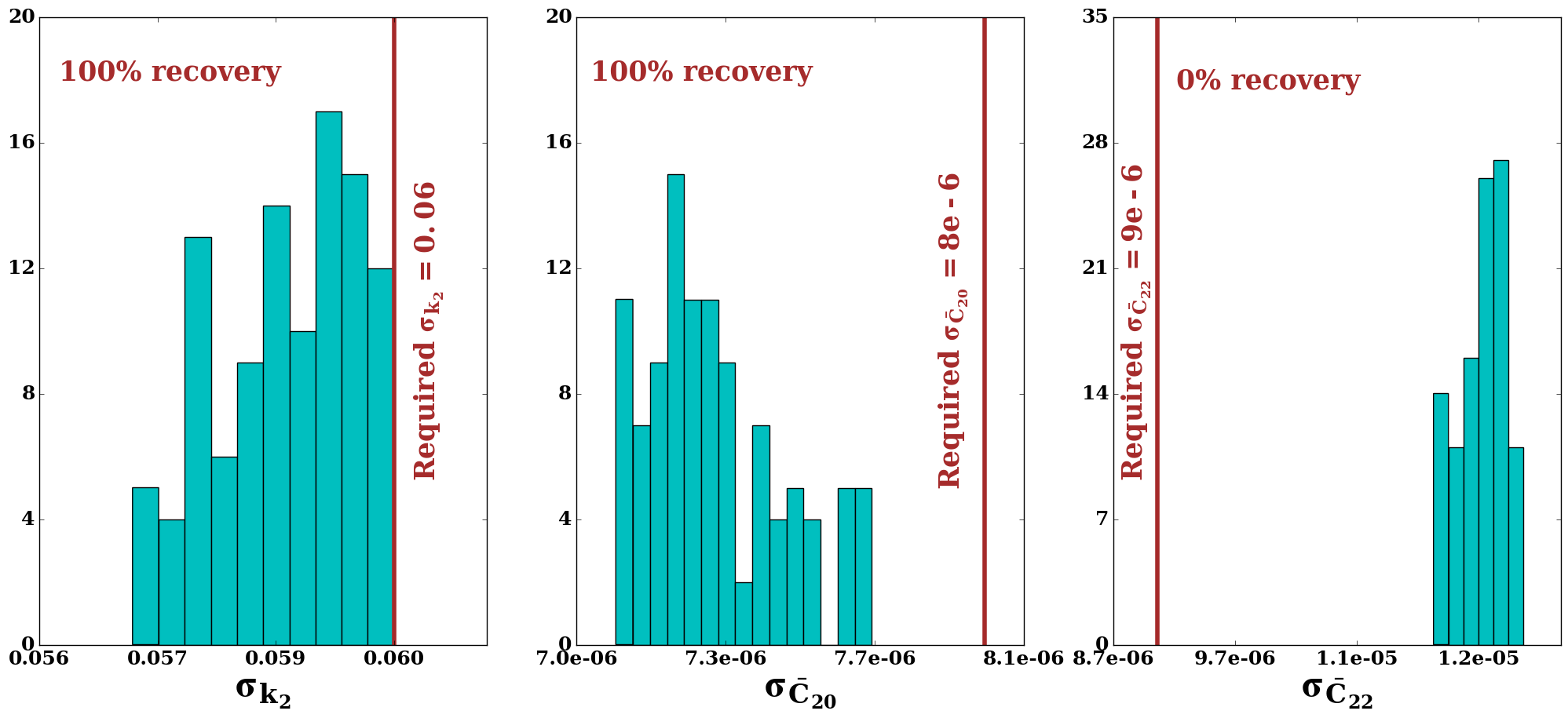}
\caption{Histograms of $k_2$, $\bar{C}_{20}$, and $\bar{C}_{22}$
  uncertainties obtained by performing covariance analyses for all
  possible combinations of 25 flybys using the
  DSN$_\textrm{2$\times$34m}$ antenna configuration with the
  latitudinal distribution shown in Table \ref{tab-preferred2x34m}
  (Scenario 2, Case Study 3).}
\label{fig-preferred2x34m}  
\end{figure*}

We found that it is possible to meet the $k_2$ measurement objective
with 25 flybys (Table \ref{tab-preferred2x34m}, Figure
\ref{fig-preferred2x34m}), as long as they include all 13 low-latitude
flybys, at least 8 mid-latitude flybys, and at least 4 high-latitude
flybys.  This result applies to the 17F12v2 trajectory and
DSN$_\textrm{2$\times$34m}$ antenna configuration.  The measurement
objective for $\bar{C}_{20}$ is also met, but that of $\bar{C}_{22}$
is never met for any combination of 25 flybys.  Failure to track a
single low-latitude flyby from the 13 available in 17F12v2 would
result in a failure to meet Clipper's primary gravity science
objective.  This fact highlights an element of risk associated with
relying on the DSN$_\textrm{2$\times$34m}$ antenna configuration.
This risk is reduced when using 70~m antennas.

 \begin{table*}[!htbp]
 \centering
 \caption{ Number of methodically selected 17F12v2 mid- and
   high-latitude flybys required to meet the tidal Love number $k_2$
   measurement objective when flybys are tracked with the
   DSN$_\textrm{2$\times$34m}$ antenna configuration and all
   low-latitude flybys are tracked (Scenario 2, Case Study 3).}
 \begin{tabular*}{\textwidth}{l@{\extracolsep{\fill}} c c c }
\hline
Europa region &  Latitude range & Avail. flybys & Req. flybys      \\ 
\hline
High latitude north  &    90$^\circ$ $\textendash$  45$^\circ$  & 7&  2\\
Mid latitude north    &   45$^\circ$ $\textendash$  15$^\circ$  & 4& 4\\
Low latitude          &   15$^\circ$ $\textendash$ -15$^\circ$  & 13 &13\\
Mid latitude south    &  -15$^\circ$ $\textendash$ -45$^\circ$  &5 & 4\\
High latitude south  &   -45$^\circ$ $\textendash$ -90$^\circ$  &2 & 2\\
\hline
Total                & 90$^\circ$ $\textendash$ -90$^\circ$& 31 & 25\\
\hline
\label{tab-preferred2x34m}
\end{tabular*}
\end{table*}

\section{Other trajectories}
\label{15N16Traj}

In an attempt to generalize our results, we examined the
suitability of the other trajectories available to us, 15F10 and
16F11, for meeting the $k_2$ measurement objective.

\begin{figure*}
\centering
\noindent
\includegraphics[width=19pc]{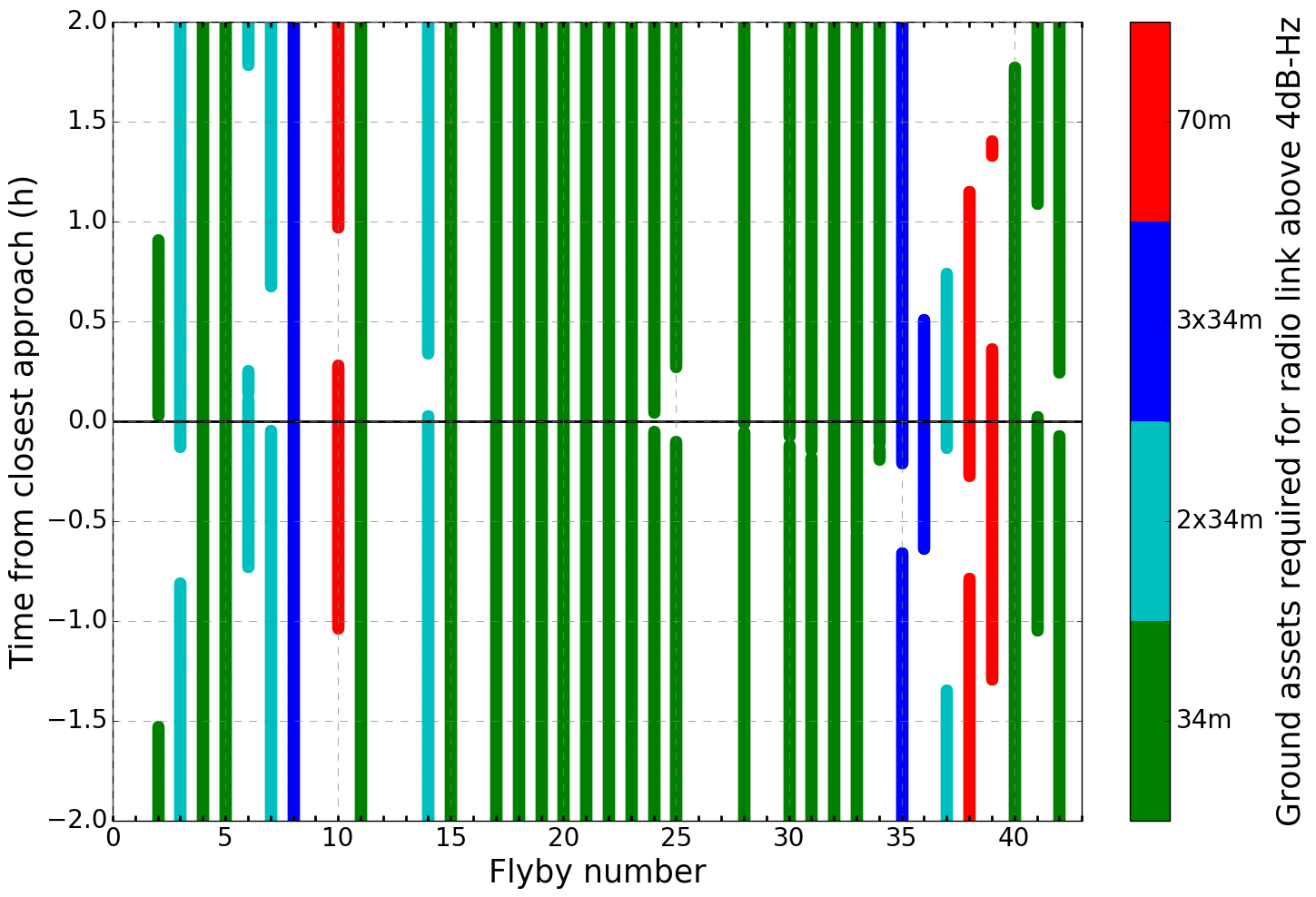}{a}
\includegraphics[width=19pc]{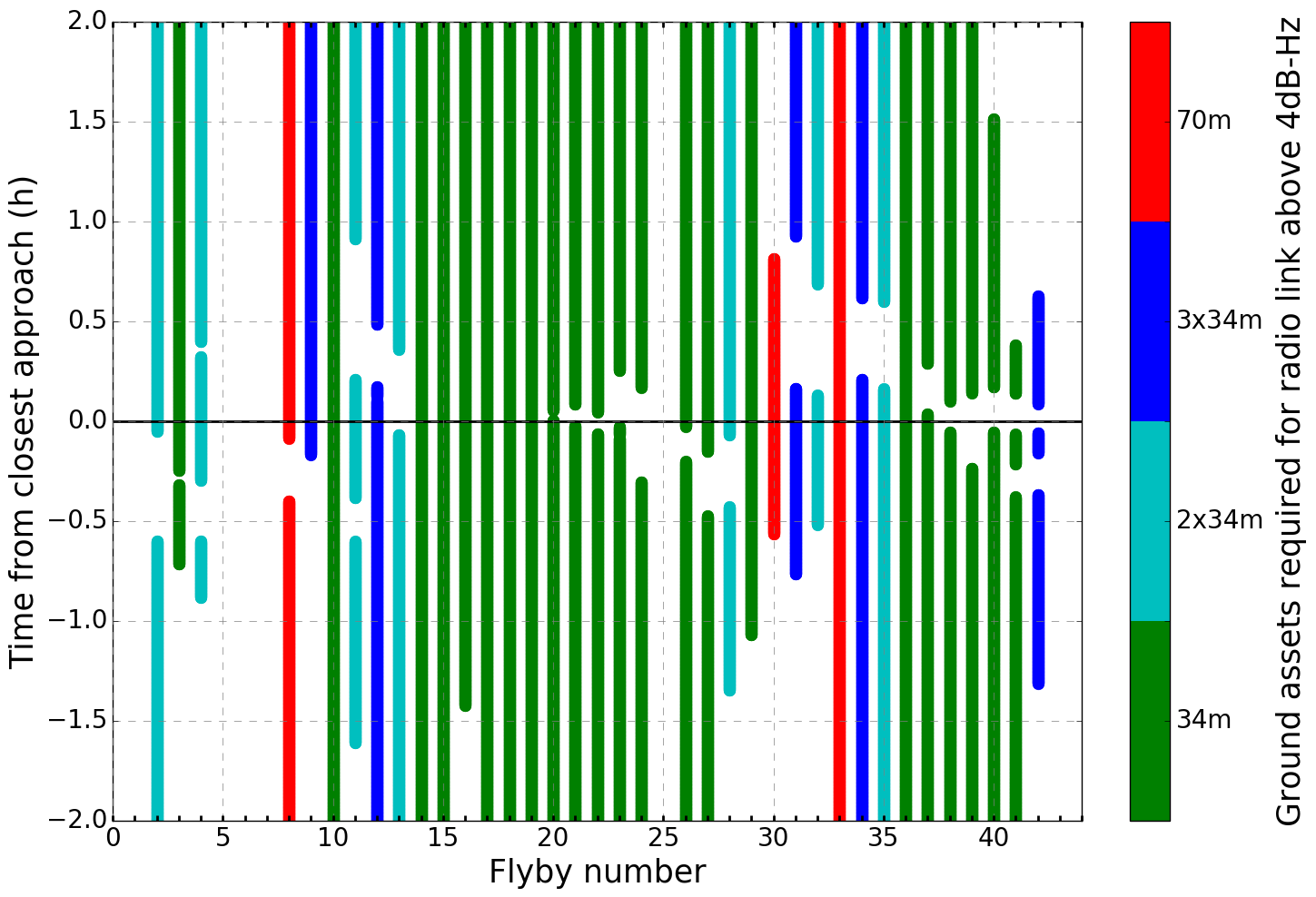}{b}
\caption{Time intervals during which a 4 dB-Hz radio link can be
  maintained for at least 1 h with trajectories 15F10 (left) and 16F11
  (right).  For each flyby, the least sensitive antenna configuration
  was used, resulting in the following 15F10 configuration: 23 tracks
  with a 34 m antenna (green), 5 tracks with a 2x34 m array (cyan), 3
  tracks with a 3x34 m array, and 3 tracks with a 70 m antenna (red)
  for 15F10.  For 16F11, the configuration includes 22 tracks with a
  34 m antenna (green), 7 tracks with a 2x34 m array (cyan), 5 tracks
  with a 3x34 m array (blue), and 3 tracks with a 70 m antenna (red).
  Flybys with low SEP angle ($<$20$^\circ$) or high altitude ($>$100
  km) were discarded.}
\label{antGain15N16}  
\end{figure*}

The 15F10 and 16F11 trajectories consist of 42 and 43 flybys,
respectively.  They yield 88 and 106 illuminated crossover points
below 1000 km altitude, respectively.  Similar to Scenario 2 of
trajectory 17F12v2 (section \ref{parTrk}), Doppler observations were
simulated with the least sensitive DSN configuration that maintains
the 4dB-Hz radio link budget for track durations of at least 1
h. Figure \ref{antGain15N16} shows the time intervals for which such a
link can be maintained with a variety of DSN assets within $\pm$2 h
of closest approach. After further discarding flybys with SEP angle
$<$ 20$^\circ$ and closest approach altitude $>$ 100~km,
a total of 34 and 37 tracked flybys remain with the 15F10 and 16F11
trajectories, respectively.  The numbers of Doppler and crossover
measurements that can be obtained with the available flybys are
shown in Table \ref{tab-15N16Msr}.

\begin{table*}
\centering
\caption{ Number of Doppler and crossover measurements that can be
  obtained with trajectories 15F10 and 16F11 with various DSN
  configurations.  The number of crossovers corresponds to the number 
  of intersections of tracked flybys. 
  The number of tracked flybys are shown in columns 2 and 5.
Also shown are the total durations (columns 4 and 7) during which
tracking can be conducted with a link budget above 4 dB-Hz, expressed
as a fraction of the total potential tracking time ($\pm$2 h of each
closest approach).  Flybys with $<$ 1 h tracking duration, SEP angle $<$
20$^\circ$, and altitude $>$ 100 km were discarded.
}
\begin{tabular*}{\textwidth}{l@{\extracolsep{\fill}} cccccccc} 
\hline
\multirow{3}{*}{DSN } & \multicolumn{4}{c}{15F10}                                    & \multicolumn{4}{c}{16F10}                \\ \cline{2-5} \cline{6-9}
                           & Tracked                    & Doppler & Crossover    & Tracking    &  Tracked & Doppler & Crossover    & Tracking   \\
                  config.  & \multicolumn{1}{c}{flybys} & msr.    & msr.         & fraction    &  flybys  & msr.    & msr.         & fraction \\ \hline
DSN$_\textrm{34m}$                   &  23              & 5065    & 35          & 50\%       &  22               & 4808      & 30        & 47\%         \\
DSN$_\textrm{2$\times$34m}$   &  23+5          & 5844    & 46          & 58\%       &  22+7           & 6052     & 59         & 59\%         \\
DSN$_\textrm{3$\times$34m}$   &  23+5+3      & 6367    & 51          & 63\%       &  22+7+5       & 6834      & 74        & 66\%         \\
DSN$_\textrm{70m}$                   &  23+5+3+3  & 6774    & 68         & 67\%       &  22+7+5+3  & 7377       & 85      & 71\%       \\ \hline
\label{tab-15N16Msr}  
\end{tabular*}
\end{table*}

\begin{table}[!htbp]
\centering
\caption{ Estimated uncertainties in tidal Love number $k_2$ and
  low-order gravity field coefficients for trajectories 15F10 and
  16F11 when 28 and 29 flybys, respectively, are tracked (duration
  $>$1 h) with the DSN$_\textrm{2$\times$34m}$ antenna configuration.
  Entries in bold indicate that the requirement (rightmost column) was
  not met.}  
\label{tab-comparison}
\begin{tabular}{c c c c} \hline
\multirow{2}{*}{Parameters} & \multicolumn{2}{c}{DSN$_\textrm{2$\times$34m}$ } & \multirow{2}{*}{Requirement} \\ \cline{2-3}
                           & 15F10        & 16F11       &                              \\ \hline
$k_2$                            &             0.053                 &    0.055                                & $\textless$0.06                \\
$\bar{C}_{20}$              &   6.0$\times$10$^{-6}$   &      {6.6$\times$10$^{-6}$} & $\textless${8.0$\times$10$^{-6}$} \\
$\bar{C}_{22}$              &  \bf{10$\times$10$^{-6}$} &  \bf{11$\times$10$^{-6}$} & $\textless${9.0$\times$10$^{-6}$} \\
$\bar{C}_{30}$              &  \bf{30$\times$10$^{-7}$} &  \bf{29$\times$10$^{-7}$} & $\textless${4.0$\times$10$^{-7}$} \\
$\bar{C}_{40}$              &  \bf{31$\times$10$^{-7}$} &  \bf{27$\times$10$^{-7}$} & $\textless${4.0$\times$10$^{-7}$} \\ \hline
\end{tabular}
\end{table}

\begin{figure*}
\centering
\noindent
\includegraphics[width=35pc]{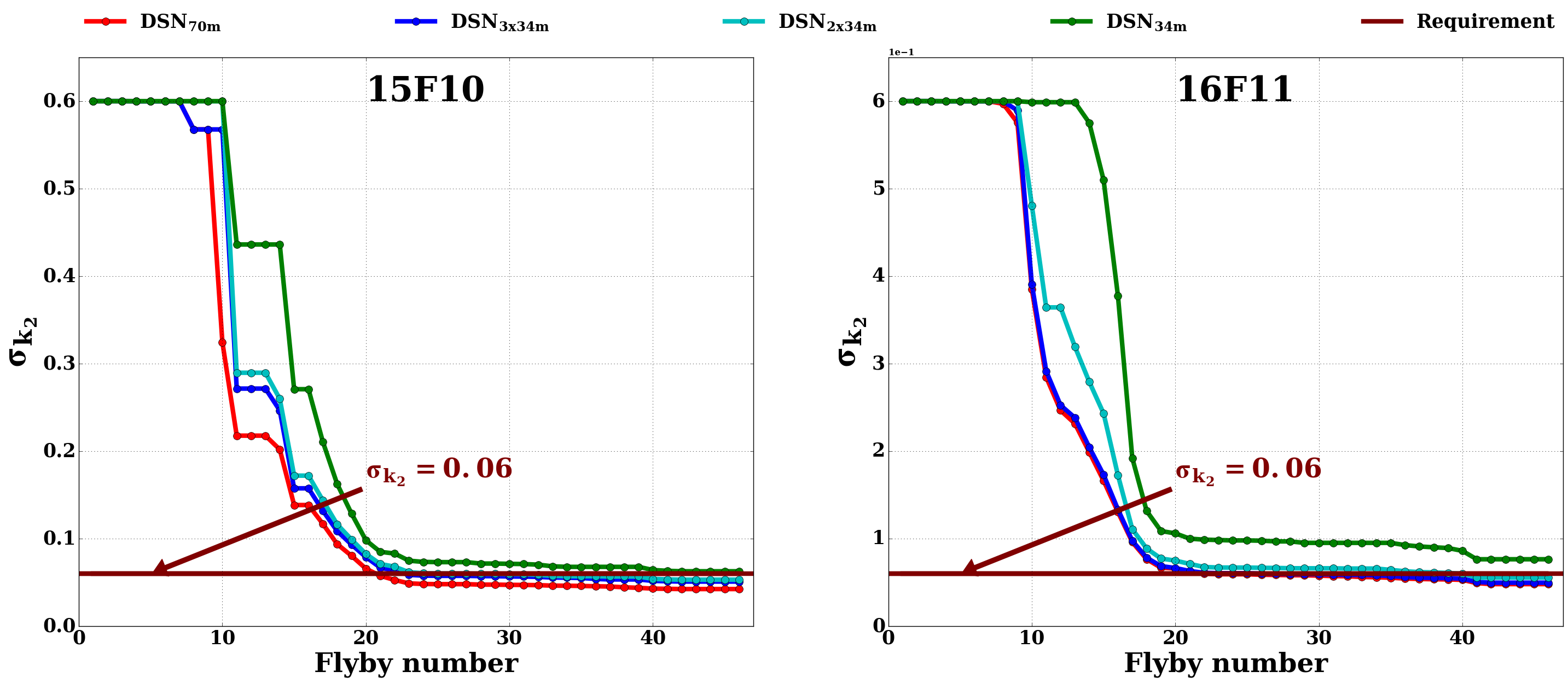}
\caption{Precision of the tidal Love number $k_2$ when data from 42(43) consecutive flybys of 15F10(16F11) trajectory are analyzed 
(flybys with $<$ 1 h tracking duration, SEP angle $<$ 20$^\circ$, and altitude $>$ 100 km were discarded) 
using progressively more sensitive DSN configurations (Figure \ref{antGain15N16}). 
The green curve shows the performance with a single 34 m antenna using a total of 23 (22) flybys (DSN$_\textrm{34m}$). 
The cyan curve considers the addition of a 2x34 m antenna array for 5 (7) flybys, a total of 28 (29) flybys (DSN$_\textrm{2$\times$34m}$). 
The blue curve considers the addition of a 3x34 m antenna for 3 (5) flybys, a total of 31 (34) flybys (DSN$_\textrm{3$\times$34m}$).  
The red curve considers the addition of a single 70 m antenna for 3 (3) flybys, a total of 34 (37) flybys (DSN$_\textrm{2$\times$34m}$)
and is nearly indistinguishable from the blue curve in the case of the 16F11 trajectory. 
The brown horizontal lines indicate the measurement objectives specified in Table \ref{tab-gswg}.}

\label{fig-comparison}  
\end{figure*}

Similar to the first case study in Scenario 2 (section
\ref{inv1-2x34}), we examined the precision of the Love number $k_2$
as data from consecutive flybys becomes available. We found that the
$k_2$ measurement objective is not achievable with a single 34 m
antenna (Figure \ref{fig-comparison}). However, as in the situation
with the 17F12v2 trajectory, the $k_2$ measurement objective can be
met (Table \ref{tab-comparison}) with both 15F10 and 16F11
trajectories and DSN$_\textrm{2$\times$34m}$ antenna configurations,
where the majority of flybys are tracked with a single 34 m antenna
and a few additional flybys are tracked with a 2x34 m antenna array
(Table \ref{tab-15N16Msr}).

We also examined the minimum number and latitudinal distribution
of tracked flybys that are required to meet the $k_2$ measurement
objective.  We selected the DSN$_\textrm{2$\times$34m}$ antenna
configuration (Table \ref{tab-15N16Msr}) to perform this case study
with 28 and 29 tracked flybys for trajectories 15F10 and 16F11,
respectively.  We selected the flybys as in Case Study 3 of Scenario 2
(section \ref{inv3-2x34}).

\begin{table*}[!htbp]
\centering
\caption{Minimum number of flybys required in each latitude band to meet the
  tidal Love number $k_2$ measurement objective with the 15F10 and 16F11 trajectories 
  when flybys are tracked with the DSN$_\textrm{2$\times$34m}$ antenna configuration.}
\begin{tabular*}{\textwidth}{l@{\extracolsep{\fill}} ccccc} 
\hline
\multirow{2}{*}{Europa region} & \multirow{2}{*}{Latitude range} & \multicolumn{2}{c}{15F10}          & \multicolumn{2}{c}{16F11}          \\ \cline{3-4} \cline{5-6}
                               &                                 & Avail. flybys & Req. flybys & Avail. flybys & Req. flybys \\ \hline
                               
High latitude north  &    90$^\circ$ $\textendash$  45$^\circ$  & 7 &  3 & 4&  2\\
Mid latitude north    &   45$^\circ$ $\textendash$  15$^\circ$  & 5 & 5 & 7& 7\\
Low latitude          &   15$^\circ$ $\textendash$ -15$^\circ$    & 11 &11 & 12 &12\\
Mid latitude south    &  -15$^\circ$ $\textendash$ -45$^\circ$  &3 & 3 &4 & 1\\
High latitude south  &   -45$^\circ$ $\textendash$ -90$^\circ$  &2 & 2 &2 & 2\\
\hline
Total                & 90$^\circ$ $\textendash$ -90$^\circ$& 28 & 24 & 29 & 24\\
\hline
\label{tab-preferred16F11}
\end{tabular*}
\end{table*}

We found that a minimum of 24 flybys with a specific latitudinal
distribution (Table \ref{tab-preferred16F11}) are sufficient to meet
the $k_2$ measurement objective for both trajectories.
As with the 17F12v2 trajectory, the measurement objective for
$\bar{C}_{20}$ is also met with the same distribution of flybys,
whereas the measurement objective for $\bar{C}_{22}$ is never met for
any combination of 24 flybys.

In summary, it takes at least 24 carefully selected flybys in 15F10
and 16F11 (Table \ref{tab-preferred16F11}) and at least 25 carefully
selected flybys in 17F12v2 (Table \ref{tab-preferred2x34m}) to meet
the $k_2$ objective with the DSN$_\textrm{2$\times$34m}$ antenna
configuration.  The similarity in the required number of flybys
suggests that it may be possible to generalize the results to other
trajectories that are similar in character (section \ref{genDist}).

\section {Generalized tracking requirements}
\label {genDist}

Here, we examine whether results that apply to trajectories 15F10,
16F11, and 17F12v2 are sufficiently similar that they can be
extrapolated to other trajectories that are similar in character.  Our
results are summarized in Table \ref{tab-4.5m}.  Taking the maximum
values of these results as a basis for extrapolation, we find that a
minimum of $\sim$13 low-latitude, $\sim$8 mid-latitude, and $\sim$5
high-latitude flybys are necessary to meet the $k_2$ objective.
Likewise, we find that a total tracking duration that is at least 50\%
of the total potential tracking time ($\pm$2 h of each closest
approach) is necessary. Expressed as a fraction of total potential
tracking time for selected flybys only (tracking duration $>$ 1 h,
altitude $<$ 100 km, SEP angle $>$ 20$^\circ$ deg), the required
percentage is 88\%.  The availability of at least 52 crossover points
in illuminated regions completes the requirements.

\begin{table*}
\centering
\caption{Summary of results obtained with 15F10, 16F11, and 17F12v2
  trajectories and the DSN$_\textrm{2$\times$34m}$ antenna
  configuration.  Tracking fraction refers to the total duration
  during which tracking can be conducted with a link budget above 4
  dB-Hz, expressed as a fraction of the total potential tracking time
  ($\pm$2 h of each closest approach).  
Flybys with $<$ 1 h tracking duration, SEP angle $<$ 20$^\circ$, and altitude $>$ 100 km were discarded.}
\begin{tabular*}{\textwidth}{c@{\extracolsep{\fill}}ccc|ccc|ccc} \hline
\multirow{4}{*}{Europa Region} &  \multicolumn{9}{c}{Trajectory}                                               \\ \cline {2-10}
                        &                           \multicolumn{3}{c}{15F10}            & \multicolumn{3}{c}{16F11}          & \multicolumn{3}{c}{17F12v2}                             \\ \cline {2-10}
\multirow{1}{*}{} & No. of & Tracking   & No. of    & No. of & Tracking   & No. of     & No. of     & Tracking       & No. of         \\
                        &                            flybys   & fraction & crossovers & flybys   & fraction & crossovers & flybys       & fraction     & crossovers     \\ \cline {1-10}
High latitude: 90$^\circ$ $\textendash$  45$^\circ$ & 5   &            &         &   4       &             &              & 4        &               &               \\
Mid latitude: 15$^\circ$ $\textendash$  15$^\circ$   & 8   &  50\%  &  38   &  8       &    49\%  &     50    &  8       &    47\%  &       52          \\
Low latitude: 15$^\circ$ $\textendash$  0$^\circ$    & 11  &            &         &  12       &             &            & 13       &               &                  \\ \hline
Total                &                                                           24 &   50\% &  38    &  24      &    49 \% &     50   & 25      &      47\%  &    52     \\ \hline         
\label{tab-4.5m}
\end{tabular*}
\end{table*}

\section{Conclusions}
\label{conclusion}
A Europa Clipper gravity science investigation can address important
mission objectives, such as confirming the presence of an ocean,
determining Europa's gravity field, quantifying the time-varying tidal
potential, verifying whether the ice shell is hydrostatic, and
providing high-precision reconstructed trajectories that other
instrument teams will greatly benefit from.

We performed covariance analyses to quantify the precision with which
geophysical parameters can be determined with a radio science
investigation and a nominal mission profile with trajectory 17F12v2.
We found that the availability of crossover measurements allows
measurement objectives to be achieved with substantially fewer tracked
flybys than in a Doppler-only scenario.  Even with 70~m antennas, the
measurement objective for the second degree and order gravitational
harmonic cannot be achieved without crossover measurements.

By simulating hundreds of thousands of combinations of tracked flybys,
we were able to quantify the distribution of flybys with
sub-spacecraft latitudes at closest approach within certain latitude
regions that provides the best prospects for meeting measurement
objectives.  We found that tracking a dozen low-latitude flybys and a
dozen mid- to high-latitude flybys are both essential.

We found that it is not possible to maintain a 4 dB-Hz radio link
budget during a $\pm$2 hour interval centered on each flyby's closest
approach epoch, even with the most sensitive ground-based assets of
the DSN.  However, we found that 45 out of 46 flybys can be tracked
for a total duration of at least one hour with a 70~m antenna.  With a
34 m antenna, 26 out of 46 flybys can be tracked for a total duration
of at least one hour, and 5 additional flybys can be tracked for a
total duration of at least one hour with a two-antenna array of 34~m
diameter antennas.  

If 70 m antennas are used, the tidal Love number $k_2$ measurement
objective can be met by tracking at least 23 methodically selected
flybys, with good resilience in case certain flybys are unexpectedly
missed.  If 34~m antennas are used without arraying, the $k_2$
measurement objective is not achievable.  However, it is achievable by
tracking 26 flybys with 34 m antennas and 5 additional flybys with
arrays of two 34~m antennas.  If flybys are carefully selected with
respect to the latitudinal distribution of closest approaches, the
$k_2$ objective can be met by tracking at least 25 flybys with 34~m
antennas and two-antenna arrays, provided that all low-latitude flybys
are tracked.  Such a configuration provides little margin for error.

By comparing our 17F12v2 results to 15F10 and 16F11 results, we showed
that our conclusions are roughly generalizable to trajectories that
are similar in character.

\section*{Acknowledgments}
AKV and JLM were supported in part by NASA Europa
Mission Project Subcontract 1569162.  This work was enabled in part by
the Mission Operations and Navigation Toolkit Environment (MONTE) and
the Spacecraft, Planet, Instrument, Camera-matrix, Events (SPICE)
toolkit.  MONTE and SPICE are developed at the Jet Propulsion
Laboratory, which is operated by Caltech under contract with NASA.  
The Europa Clipper trajectory data used in this work is available from the NAIF server at
\url{ftp://naif.jpl.nasa.gov/pub/naif/EUROPACLIPPER}. We
thank James Roberts and Robert Pappalardo with assistance in
specifying gravity science measurement requirements.  We thank Dipak
Srinivasan, Peter Ilott, Avinash Sharma, and Ryan Park with assistance
in providing values of system parameters.  We thank Gregor
Steinbr\"ugge for providing estimates of range uncertainties at
crossover locations. We thank Francis Nimmo for providing helpful
comments on the manuscript and GSWG members for informative
discussions.








\section*{References}
\bibliography{reference}
\end{document}